\documentclass[11pt]{article}

\usepackage[final]{acl}
\usepackage{kotex}
\usepackage{booktabs} 
\usepackage{array}    
\usepackage{multirow} 
\usepackage{times}
\usepackage{latexsym}
\usepackage{subcaption}
\usepackage{microtype}
\usepackage[T1]{fontenc}
\usepackage{tipa}
\usepackage[utf8]{inputenc}

\usepackage{microtype}

\usepackage{inconsolata}

\usepackage{graphicx}
\usepackage{soul}
\usepackage[linguistics]{forest}
\usepackage{latexsym}
\usepackage{amssymb}
\usepackage{amsmath}
\usepackage{tikz}
\usepackage{float}
\usepackage[utf8]{inputenc}
\usepackage{microtype}
\usepackage{tipa}
\usepackage{inconsolata}
\usepackage{palatino}
\usepackage{stmaryrd}
\graphicspath{{./images/}}
\usepackage[dvipsnames]{xcolor}
\usepackage{setspace}
\usepackage{adjustbox}
\usepackage{enumitem}
\usetikzlibrary{positioning, tikzmark}
\usepackage{tikz-qtree}
\usepackage[varwidth]{ragged2e}
\usepackage{multirow}
\usepackage{titling}
\usepackage{times}
\usepackage{fancyhdr}
\setkeys{Gin}{keepaspectratio}
\usepackage{caption}
\usepackage{setspace} 
\usepackage{titlesec} 

\title{Deep Supervised Contrastive Learning of Pitch Contours for Robust Pitch Accent Classification in Seoul Korean}

\author{
  \textbf{Hyunjung Joo}\textsuperscript{1,3}, 
  \textbf{GyeongTaek Lee}\textsuperscript{2}\thanks{~Corresponding author.} \\
  \textsuperscript{1}Department of Linguistics, Rutgers University\\
  \textsuperscript{2}Department of Smart Factory, Gachon University \\
  \textsuperscript{3}Hanyang Institute for Phonetics and Cognitive Sciences of Language (HIPCS) \\
  \texttt{hyunjung.joo@rutgers.edu}, \texttt{leegt@gachon.ac.kr}
}
\begin{document}
\maketitle

\begin{abstract}

The intonational structure of Seoul Korean has been defined with discrete tonal categories within the Autosegmental-Metrical model of intonational phonology. However, it is challenging to map continuous $F_0$ contours to these invariant categories due to variable $F_0$ realizations in real-world speech. Our paper proposes Dual-Glob, a deep supervised contrastive learning framework to robustly classify fine-grained pitch accent patterns in Seoul Korean. Unlike conventional local predictive models, our approach captures holistic $F_0$ contour shapes by enforcing structural consistency between clean and augmented views in a shared latent space. To this aim, we introduce the first large-scale benchmark dataset, consisting of manually annotated 10,093 Accentual Phrases in Seoul Korean. Experimental results show that our Dual-Glob significantly outperforms strong baseline models with state-of-the-art accuracy (77.75\%) and F1-score (51.54\%). Therefore, our work supports AM-based intonational phonology using data-driven methodology, showing that deep contrastive learning effectively captures holistic structural features of continuous $F_0$ contours.

\end{abstract}

\section{Introduction}

Seoul Korean is an edge-prominence language (\citealp{jun1998accentual,jun2005korean}), where fundamental frequency $F_0$, the acoustic correlate of pitch, is used to organize prosodic structures to encode grammatical and pragmatic distinctions. Within the Autosegmental-Metrical (AM) theory of intonation (e.g., \citealp{beckman1986intonational}; \citealp{ladd2008intonational}) and its Korean Tones and Break Indices (K-ToBI; \citealp{jun2000k}) transcription system, a continuous $F_0$ contour is modeled as a sequence of discrete tonal targets such as Lows (L) and Highs (H), which are interpolated with one another.

\begin{figure}[t]
\centering
\begin{tikzpicture}[
    baseline=(IP.north),
    every node/.style={font=\small, inner sep=2pt},
    level 1/.style={sibling distance=80pt, level distance=22pt},
    level 2/.style={sibling distance=24pt, level distance=22pt},
    level 3/.style={sibling distance=10pt, level distance=22pt},
    level 4/.style={level distance=20pt, level distance=17pt},
    edge from parent/.style={draw, -}
]

\node (IP) {Intonational Phrase (IP)}
    child { node {Accentual Phrase (AP)}
        child { node {Word (W)}
            child { node {$\sigma$} 
                child { node (T1) {T} }
            }
        }
        child { node {} edge from parent[draw=none]
            child { node {$\sigma$} edge from parent[draw=none]
                child { node (T1) {H} }
            }
        }
        child { node {(W)} edge from parent[draw=none]
            child { node {$\dots$} edge from parent[draw=none]
            }
        }
        child { node {} edge from parent[draw=none]
            child { node {$\sigma$} edge from parent[draw=none]
                child { node (T1) {L} }
            }
        }
        child { node {(W)}
            child { node (Sn) {$\sigma$} 
                child { node (Tn) {H} }
            }
        }
    }
    child { node {(AP)} };

\node[right=15pt of Tn, yshift=0pt] (BT) {\%};
\draw (Sn.south) -- (BT);
\draw (BT.north) .. controls +(2,1) and +(2,0.3) .. (IP.south);
\end{tikzpicture}
\caption{Intonational structure of Seoul Korean \cite{jun1998accentual}. The AP-initial tone (T) is realized as H for aspirated and tense consonants, otherwise L. The \% symbol refers to a boundary tone (e.g., L\% or H\%) at the end of an IP.}
\label{intonationalstructure}
\vspace{-18pt}
\end{figure}
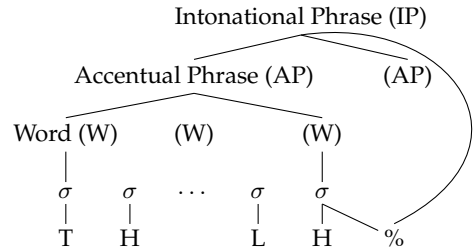

The intonational structure of Seoul Korean is hierarchically organized, with an Accentual Phrase (AP) as the basic unit, one or more of which are grouped into an Intonational Phrase (IP). According to \citet{jun1998accentual}, APs with more than three syllables typically surface as LHLH or HHLH, depending on the phrase-initial segment: APs beginning with aspirated or tense consonants surface as HHLH, while others surface as LHLH (\autoref{intonationalstructure}). Shorter APs show fourteen possible tonal patterns: LH, HH, LL, HL, LLH, LHH, HLH, LHL, HHL, HLL, LHLL, HHLL, LHLH, and HHLH (See \textbf{Appendix A} for the schematic contours of these patterns.). 

Despite this well-established theoretical characterization of intonation, there is a significant gap between phonological models \cite{jun1998accentual} and real-world acoustic data. Crucially, it is highly complicated to map continuous $F_0$ contours onto discrete and invariant tonal categories due to the inherent variability of $F_0$ coming from gender differences, speech styles, and phonetic contexts (e.g., \citealp{cole2016new}). While intonational research has been conducted based on expert annotations using the ToBI system \cite{beckman1994tobi}, such perception-based transcriptions may fall into subjective bias and are difficult to deal with large-scale data for computational modeling.

To fill this gap, we propose a \textbf{Dual-Glob}, a deep supervised contrastive learning approach to characterize continuous $F_0$ variations into fine-grained and invariant pitch accent categories in Seoul Korean. Unlike traditional supervised models that are prone to overfitting and sensitive to measurement noise, our contrastive framework enables the model to learn robust and discriminative representations of raw $F_0$ contours. Therefore, we maximized the similarity between $F_0$ contour shapes within the same pitch accent category while contrasting them across different pitch accent categories in the latent space.

This study supports the discrete tonal categories defined within AM theory with the representional power of data-driven deep learning approaches. By using large-scale acoustic data, we show that our model effectively captures fine-grained pitch accent categories in Seoul Korean. Our proposed framework offers a more robust and scalable approach to classifying the pitch accent patterns than conventional models.

The contributions of our work are summarized as follows:
\vspace{-5pt}
\begin{itemize}
    \item \textbf{Large-Scale Prosodic Benchmark Dataset:} We construct the first large-scale benchmark dataset for classifying pitch accent categories in Seoul Korean, paving the way for future intonational research.
\vspace{-18pt}
    \item \textbf{Connecting AM Theory with Deep Learning Framework:} 
    To our knowledge, we propose the first deep learning model that account for the AM's discrete tonal representation with holistic $F_0$ contours. This approach enriches AM-based intonational phonology with continuous and data-driven representation learning.
\vspace{-18pt}
    \item \textbf{State-of-the-Art Performance:} Experimental results show that our \textbf{Dual-Glob} method significantly outperforms other competitive baseline models with state-of-the-art accuracy in characterizing continous $F_0$ contours.
\end{itemize}

\section{Related Work}

\paragraph{Intonational Structure of  Seoul Korean}
The intonational structure of Seoul Korean has been mainly formalized within the AM framework (e.g., \citealp{beckman1986intonational,ladd2008intonational}), with an AP as the basic unit of tonal organization. (\citealp{jun1998accentual}, \citealp{jun2005korean}). \cite{jun2000k} identified up to fourteen AP tonal patterns (e.g., HHLH, LH), conditioned by the laryngeal features of the phrase-initial segment and syllable count. Studies have empirically shown that the tonal patterns of APs in Seoul Korean exhibit strong edge-prominent characteristics (e.g., \citealp{hatcher2024focus, kim2008intonational}). Importantly, \citet{kim2008intonational} found that in both read and radio speech, most APs begin with a rising tone and end with a high tone (LH...LH) to signal prosodic edges for word segmentation. Despite this basic pattern, other tonal patterns were distributed variably across registers: radio speech showed a more varied distribution than read speech.

While AM theory has been a mainstream theoretical framework in intonational phonology, configurational approaches (e.g., \citealp{bolinger1951intonation,hart2003perceptual,xu2005speech}) view intonation as a holistic $F_0$ contour shape rather than a sequence of discrete tonal targets. Interestingly, recent studies have emphasized the importance of incorporating continuous $F_0$ information (\citealp{barnes2012tonal,barnes2021and,joo2025perception}), suggesting that only considering discrete tonal targets within AM theory may be insufficient to fully capture the intonational patterns.

\paragraph{Intonational Modelling}
Inspired by configurational approaches, \citet{levow2005context} used uniform representations of pitch, duration, and intensity within a support vector machine to classify Mandarin tones and English pitch accents, while also considering the preceding and following shape contexts. However, these methods only focused on discrete features, which are not enough to capture fine-grained dynamic patterns of $F_0$. 

Recent work has modeled intonation as a continuous $F_0$ contour. The dynamical systems approach models English pitch accents as nonlinear trajectories toward phonological targets (\citealp{iskarous2017relation,iskarous2023american,iskarous2026quantal}). However, it uses predefined parameters (e.g., $F_0$ peak and velocity) in a differential equation, rather than learning representations directly from raw $F_0$ time series.

In contrast, recent deep learning approaches have shown that directly modeling raw $F_0$ of Mandarin tones can outperform conventional feature-based methods \cite{chen2022computational}. 

\paragraph{Deep Learning Approaches in Korean Prosody.}
Recent deep learning approaches use $F_0$ contours for tasks like speech emotion recognition using dual recurrent encoder~\cite{yoon2018multimodal}, or dialect identification via Bidirectional long short-term memory (BiLSTM)~\cite{lee2021korean}. However, these works largely focus on boundary detection or broad regional categories. To our knowledge, we are the first to apply deep contrastive learning to the fine-grained classification of tonal patterns in Seoul Korean (e.g., LHLH vs. HH), explicitly modeling holistic $F_0$ contour shapes.

\paragraph{Contrastive Learning for Robust Representations.}
While models like InceptionTime~\cite{ismail2020inceptiontime} handle time-series, their reliance on a large amount of labeled data limits scalability. Self-supervised predictive coding~\cite{oord2018representation} offers an alternative but rests on Markovian assumptions not suitable for holistic tonal representations (e.g., HHLH). In contrast, supervised contrastive learning (SupCon)~\cite{khosla2020supervised} can optimize for the consistency of global $F_0$ contours. By clustering the same class samples in the latent space, SupCon learns invariant tonal representations robust to measurement noise and speaker variability.

\begin{figure*}[t]
    \centering
    \includegraphics[width=0.85\textwidth]{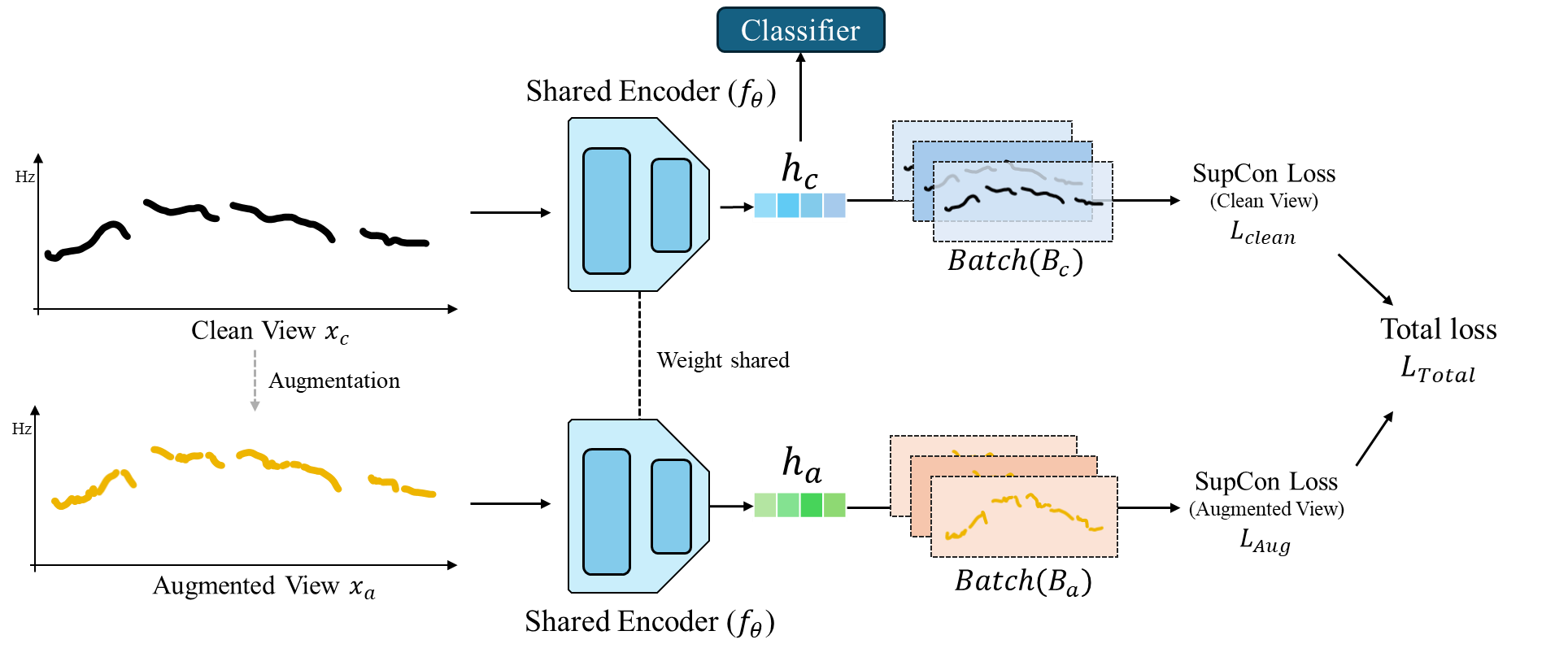} 
    \caption{Overview of the proposed \textbf{Dual-Glob} framework. The model processes entire $F_0$ contours via parallel clean ($x_c$) and augmented ($x_a$) views using a shared encoder. A composite supervised contrastive objective ($\mathcal{L}_{Total}$) enforces structural consistency across both views to learn robust representations.}
    \label{fig:dual_glob_overview}
    \vspace{-0.5cm} 
\end{figure*}

\section{Proposed Model}

\subsection{Motivation: Holistic Tonal representations vs. Temporal Prediction}
Standard self-supervised time-series methods often rely on predictive coding, predicting future segments from past contexts (e.g., $x_{past} \to x_{future}$). While this works well for stochastic processes, we argue that it is suboptimal for modeling Seoul Korean pitch accent patterns.

In AM theory, APs are not Markovian sequences but \textbf{discrete tonal representations} where the entire contour determines linguistic meaning. Thus, treating intonation as a local predictive task risks fragmenting these global units. To address this, we propose a \textbf{holistic representation learning} framework by capturing the global shape of $F_0$ contours, at the same time, making distinct representations contrastive despite local temporal variations.

\subsection{Dual-View Supervised Contrastive Learning}
To robustly learn tonal representations, we employ a dual-branch architecture (Figure \ref{fig:dual_glob_overview}) that processes both original and augmented views via shared encoders. This design enforces \textit{consistency regularization}, ensuring the model recognizes underlying tonal categories even under input perturbations.

\subsubsection{Data Augmentation and Encoding}
Given a clean input $x_c$, we generate an augmented view $x_a$ via stochastic perturbations to simulate natural variability. Both $x_c$ and $x_a$ are processed by a shared encoder $E(\cdot)$ and projection head $P(\cdot)$ to yield:
\begin{equation}
    z_c = P(E(x_c)), \quad z_a = P(E(x_a))
\end{equation}
These projections map $F_0$ contours into a normalized latent space for contrastive learning.

\subsubsection{Objective Function}
We employ the SupCon loss to structure the latent space. Specifically, we formulate a joint objective with two terms to balance intra-class compactness on clean data and robust representation learning against perturbations.

\paragraph{1. Clean-view SupCon ($\mathcal{L}_{Clean}$):}
This term ensures that the representation of the original $F_0$ contour correctly aligns with other instances of the same tonal category. Formally, the loss is defined as:

\begin{equation}
\begin{split}
    \mathcal{L}_{Clean} = - \sum_{i \in \mathcal{B}} \frac{1}{|P(i)|} \sum_{j \in P(i)} \\
    \log \frac{\exp(z_{i,c} \cdot z_{j,c} / \tau)}{\sum_{k \in \mathcal{B} \setminus \{i\}} \exp(z_{i,c} \cdot z_{k,c} / \tau)}
\end{split}
\label{eq:clean_loss}
\end{equation}
where $z_{i,c}$ is the clean projection of the $i$-th sample, $P(i)$ is the set of indices of positive samples in batch $\mathcal{B}$, and $\tau$ is a temperature parameter.

\paragraph{2. Augmented-view SupCon ($\mathcal{L}_{Aug}$):}
This term addresses the inherent instability of pitch extraction in real-world environment. 
Raw $F_0$ contours are accompanied by discontinuations (e.g., breath, plosives) and pitch tracking errors. 
Models trained solely on raw $F_0$ contours risk overfitting to surface-level irregularities. To mitigate this, $\mathcal{L}_{Aug}$ enforces invariant representation learning, enabling the model to ignore noise and capture robust phonological representations. Formally, the loss minimizes the discrepancy between augmented views and clean positives:
\vspace{-3pt}
\begin{equation}
\begin{split}
    \mathcal{L}_{Aug} = - \sum_{i \in \mathcal{B}} \frac{1}{|P(i)|} \sum_{j \in P(i)} \\
    \log \frac{\exp(z_{i,a} \cdot z_{j,c} / \tau)}{\sum_{k \in \mathcal{B} \setminus \{i\}} \exp(z_{i,a} \cdot z_{k,c} / \tau)}
\end{split}
\label{eq:aug_loss}
\end{equation}
Here, $z_{i,a}$ denotes the embedding of the augmented anchor, while $P(i)$ is the set of indices for clean samples belonging to the same class as $i$. The terms $z_{j,c}$ and $z_{k,c}$ represent these clean positive prototypes and the entire pool of clean embeddings in the batch, respectively. This asymmetric formulation explicitly encourages the model to project distorted representations onto the stable manifold formed by the clean signals, effectively denoising the intonational features.

The final training objective integrates both clean-signal consistency and augmented-view robustness. We define the total loss as a weighted sum of the two components:

\begin{equation}
    \mathcal{L}_{Total} = \lambda_{1} \mathcal{L}_{Clean} + \lambda_{2} \mathcal{L}_{Aug}
    \label{eq:total_loss}
\end{equation}
where $\lambda$ is a balance parameter. By minimizing this total loss, the model learns to capture the global $F_0$ contour shape (e.g., HHLH), ignoring small local errors that often mislead predictive models.

\subsubsection{Downstream Classification}
After training, we freeze the encoder and discard projection heads to extract latent pitch features. These are fed into standard classifiers, ensuring that performance reflects robustness of representation rather than classifier complexity.

\section{Experiments}
\subsection{Dataset and Preprocessing}
We constructed a new dataset of APs in Seoul Korean\footnote{Our dataset is available in https://github.com/hyunjungjoo/Accentual-Phrases-in-Seoul-Korean }, using the broadcasting conversational data, a large-scale corpus provided by AI Hub ~\cite{aihub2022broadcasting}. In order to make sure clear prosodic realization with precise articulation, we selected the recordings produced by 18 professional broadcasters (11 females, 7 males).

The raw audio was manually segmented into APs based on perceptual judgment and visual inspection of $F_0$ contours, since automated forced alignment often fails to detect prosodic boundaries and varying speech rates can result in different phrasing \citep{jun2003effect}.

Each AP was then annotated by two trained K-ToBI \cite{jun2000k} transcribers. A total of 10,093 APs were categorized into 16 distinct tonal patterns, including monosyllabic APs with either a L or an H tone. The distribution of these categories is provided in Table \ref{tab:dataset_dist}.

\paragraph{Feature Extraction}
We extracted $F_0$ contours using the pYIN algorithm from 22.05 kHz audio. Frame and hop lengths were set to 1024 and 256, with a range of 80--400 Hz. All sequences were fixed to 200 frames ($T=200$) for consistency.

\paragraph{Normalization}
Since absolute pitch varies by speaker (e.g., gender), raw $F_0$ introduces bias. To mitigate this, we applied speaker-wise Min-Max normalization: $x' = \frac{x - \min_k}{\max_k - \min_k}$, scaling pitch values to the range $[0, 1]$.

\begin{table}[t]
    \centering
    \footnotesize
    \renewcommand{\arraystretch}{0.85}

    \caption{Distribution of the 16 tonal labels in the dataset.}
    \label{tab:dataset_dist}
    
    \resizebox{0.8\linewidth}{!}{
    \begin{tabular}{l|r||l|r}
        \toprule
        \textbf{Label} & \textbf{Count} & \textbf{Label} & \textbf{Count} \\
        \midrule
        H      & 200   & L      & 15    \\
        HH     & 1,168 & LH     & 1,318 \\
        HHL    & 77    & LHH    & 1,084 \\
        HHLH   & 1,463 & LHL    & 81    \\
        HHLL   & 784   & LHLH   & 2,705 \\
        HL     & 57    & LHLL   & 431   \\
        HLH    & 259   & LL     & 8     \\
        HLL    & 26    & LLH    & 417   \\
        \midrule
        \textbf{Total} & \multicolumn{3}{c}{\textbf{10,093}} \\
        \bottomrule
    \end{tabular}
    }
\end{table}

\subsection{Experimental Design}
\paragraph{Baselines.}
We compared our framework against diverse competitive models. First, we employed standard sequence encoders: \textbf{1D-CNN}~\cite{wang2017time}, \textbf{BiLSTM}~\cite{schuster1997bidirectional, hochreiter1997long}, and \textbf{Transformer}~\cite{vaswani2017attention}. Second, we benchmarked against state-of-the-art time-series models: \textbf{InceptionTime}~\cite{ismail2020inceptiontime}, \textbf{TimesNet}~\cite{wu2022timesnet}, and \textbf{DLinear}~\cite{zeng2023transformers}, which specialize in trend decomposition. Finally, we included \textbf{MiniRocket}~\cite{dempster2021minirocket} as an efficient non-deep learning baseline.

\paragraph{Evaluation Protocol.}
For the proposed model, we froze the pre-trained encoder and evaluated the representations using \textbf{LightGBM}~\cite{ke2017lightgbm}, \textbf{logistic regression (LR)}~\cite{cox1958regression}, and \textbf{random forest (RF)}~\cite{breiman2001random}. All experiments were conducted using 5-fold cross-validation.
For the performance measurement, we employed both \textbf{accuracy (Acc)} and \textbf{macro-F1 score (F1)}. The detailed experimental design and the architectural specifications for our model and all comparative baselines are provided in \textbf{Appendix B}.
In addition, we employed a stochastic data augmentation strategy for contrastive learning. Detailed information regarding this procedure is provided in \textbf{Appendix C}.

\subsubsection{Ablation Study Design}
To validate the synergy between augmentation and training objectives, we evaluated representation quality using classifiers trained on frozen features. We denoted predictive loss as $\mathcal{L}_{Pred}$, with superscripts $Clean$ and $Augment$ indicating clean and augmented views. The variants were categorized as follows:

\begin{itemize}
    \item \textbf{Local Predictive Models:} To assess the impact of learning local temporal transitions, we evaluate two variants of predictive coding. 
    \textbf{Pred-C} (based on Vanilla SimTS~\cite{zheng2023simts}) employs a unidirectional contrastive loss ($\mathcal{L}_{Pred}^{Clean}$) that predicts the future half of a pitch sequence from its past, using only clean data. 
    In contrast, \textbf{Pred-A} extends this approach by utilizing bidirectional prediction between clean and augmented views ($\mathcal{L}_{Pred}^{Augment}$). Specifically, it simultaneously optimizes the prediction of the clean future from an augmented past, and vice versa, to enforce stronger local consistency across stochastic perturbations.

    \item \textbf{Global Contrastive Models:} \textbf{Glob-Clean} and \textbf{Glob-Augment} apply contrastive objectives to clean ($\mathcal{L}_{Clean}$) and augmented ($\mathcal{L}_{Aug}$) sequences, respectively, to capture holistic structural features. Each objective follows the contrastive formulation defined in \textbf{Eq. (4)}, applied to its respective data view.
 
    \item \textbf{Hybrid:} This model integrates the global contrastive objective on augmented views with a cross-view predictive task. Specifically, it combines $\mathcal{L}_{Aug}$ with a loss that forecasts the future pitch features of an augmented view from the corresponding clean future segment ($\mathcal{L}_{Aug} + \mathcal{L}_{Pred}^{Clean_{fut} \to Aug_{fut}}$). This hybrid approach encourages the model to align global structural representations while maintaining local consistency between clean and perturbed pitch contours.

\end{itemize}

\paragraph{Loss Formulation Strategies.} To justify our selection of independent loss terms for clean and augmented views ($\mathcal{L}_{Clean} + \mathcal{L}_{Aug}$), we evaluate two alternative contrastive formulations that incorporate cross-view interactions:

\begin{enumerate}
    \item \textbf{Cross-Branch Class-Aware Loss (Cross-view SupCon):} This variant explicitly calculates the contrastive loss between clean view embeddings ($z$) and augmented view embeddings ($z'$) to enforce direct cross-view alignment. The objective is defined as:
    \begin{equation}
    \begin{split}
        \mathcal{L}_{cross} = &-\frac{1}{|I|} \sum_{i \in I} \frac{1}{|P(i)|} \sum_{p \in P(i)} \\
        &\log \frac{\exp(\text{sim}(z_i, z'_p)/\tau)}{\sum_{a \in I} \exp(\text{sim}(z_i, z'_a)/\tau)}
    \end{split}
    \end{equation}
    where $I$ denotes the set of indices in a batch, and $P(i) = \{p \in I \mid y'_p = y_i\}$ is the set of indices of augmented samples sharing the same class label as clean sample $i$.
    
    \item \textbf{Unified SupCon:} This variant treats clean and augmented views as a single integrated batch. We define a combined set $Z_{all} = \{z_1, \dots, z_{|I|}, z'_1, \dots, z'_{|I|}\}$ with indices $J$. The objective is to optimize the supervised contrastive loss over this unified batch:
    \begin{equation}
    \begin{split}
        \mathcal{L}_{unified} = &-\frac{1}{|J|} \sum_{i \in J} \frac{1}{|P(i)|} \sum_{p \in P(i)} \\
        &\log \frac{\exp(\text{sim}(r_i, r_p)/\tau)}{\sum_{a \in J \setminus \{i\}} \exp(\text{sim}(r_i, r_a)/\tau)}
    \end{split}
    \end{equation}
    where $r$ denotes an embedding in $Z_{all}$ and $P(i) = \{p \in J \setminus \{i\} \mid y_p = y_i\}$.
\end{enumerate}

\paragraph{Gender-Specific Analysis.}
Previous studies (\citealp{henton1989fact, pepiot2014male}) show that $F_0$ from female speakers ranged wider and is more variable \text{than} that from male speakers. We hypothesized these different $F_0$ realizations could \text{hinder learning} the shared tonal patterns. To investigate this, we designed two settings:

\begin{enumerate}
    \item \textbf{Disaggregated Evaluation:} We evaluated the unified model on male and female subsets separately to detect potential bias.
    \item \textbf{Gender-Specific Training:} We trained independent models on gender-split data to assess if domain splitting outperforms learning a shared invariant space.
\end{enumerate}

\begin{table}[t]
    \centering
    \caption{Comparison of classification performance between the proposed Dual-Glob framework and baseline models. All results represent the average of 5-fold cross-validation (Mean $\pm$ SD).}
    \label{tab:main_results_updated}
    \resizebox{\linewidth}{!}{
    \begin{tabular}{lcc}
        \toprule
        \textbf{Model} & \textbf{Acc} & \textbf{F1} \\
        \midrule
        \multicolumn{3}{l}{\textit{Standard Deep Learning Baselines}} \\
        1D-CNN      & 0.7410 \small{$\pm$ 0.0104} & 0.4930 \small{$\pm$ 0.0134} \\
        BiLSTM      & 0.7568 \small{$\pm$ 0.0156} & 0.4915 \small{$\pm$ 0.0290} \\
        Transformer & 0.7177 \small{$\pm$ 0.0107} & 0.4680 \small{$\pm$ 0.0248} \\
        \midrule
        \multicolumn{3}{l}{\textit{State-of-the-Art Time-Series Models}} \\
        InceptionTime & 0.7426 \small{$\pm$ 0.0106} & 0.5043 \small{$\pm$ 0.0147} \\
        TimesNet      & 0.6794 \small{$\pm$ 0.0180} & 0.3759 \small{$\pm$ 0.0191} \\
        MiniRocket    & 0.7303 \small{$\pm$ 0.0152} & 0.4322 \small{$\pm$ 0.0179} \\
        DLinear       & 0.6461 \small{$\pm$ 0.0078} & 0.3892 \small{$\pm$ 0.0242} \\
        \midrule
        \textbf{Proposed (Dual-Glob)} & & \\
        \hspace{3mm} w/ LightGBM      & 0.7743 \small{$\pm$ 0.0052} & 0.5086 \small{$\pm$ 0.0064} \\
        \hspace{3mm} w/ RF            & 0.7740 \small{$\pm$ 0.0069} & 0.5051 \small{$\pm$ 0.0061} \\
        \hspace{3mm} w/ LR            & \textbf{0.7775} \small{$\pm$ 0.0064} & \textbf{0.5154} \small{$\pm$ 0.0151} \\
        \bottomrule
    \end{tabular}
    }
\end{table}

\subsection{Result Analysis}
\paragraph{Comparative Analysis with Baselines.}
Table \ref{tab:main_results_updated} shows the main result of the proposed model and baseline models. Our \textbf{Dual-Glob} (w/ LR) achieved state-of-the-art Acc (\textbf{0.7775}) and F1 (\textbf{0.5154}), outperforming the strongest baseline, \textbf{BiLSTM} (0.7568). While \textbf{BiLSTM} remained competitive, our method demonstrated superior stability with lower standard deviations, suggesting contrastive learning effectively stabilized performance. \textbf{InceptionTime} (0.7426) and \textbf{MiniRocket} (0.7303) showed strong results but fall short, while \textbf{TimesNet} (0.6794) failed to capture pitch variations.

\paragraph{Limitations of Attention and Decomposition.}
\textbf{Transformer} (0.7177) and \textbf{DLinear} (0.6461) performed poorly. DLinear's trend decomposition proves too rigid for complex intonation. Similarly, Transformer's global attention failed to capture fine-grained local transitions effectively compared to our approach. For detailed error analysis, refer to \textbf{Appendix D}.

\begin{figure}[t]
    \centering
    \includegraphics[width=0.9\linewidth]{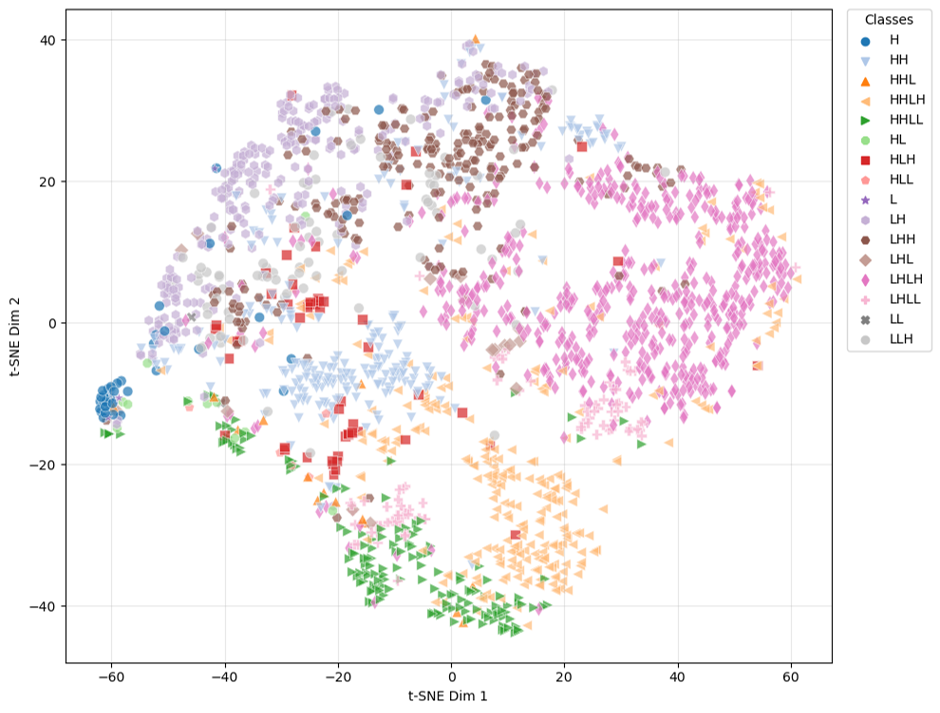}
    \caption{t-SNE visualization of the validation set. Distinct clusters form for most categories, while overlaps between similar classes (e.g., \text{HHL} vs. \text{HHLL}) reflect their $F_0$ contour shape resemblance.}
    \label{fig:all_tsne}
        \vspace{-0.25cm} 
\end{figure}

\paragraph{Feature Visualization}
Figure \ref{fig:all_tsne} visualizes representations via t-SNE~\cite{maaten2008visualizing}, showing our model maps distinct tonal patterns into clusters, capturing global pitch shapes. However, similar categories like HHL and HHLL overlap. Even though the final pitch patterns are different, the initial pitch patterns are the same so that those categories are placed closely in the latent space.

\subsection{Analysis of Ablation Studies}
\paragraph{Limitations of Predictive Contrastive Modeling.} Table \ref{tab:ablation} highlights the limitations of the predictive contrastive paradigm (SimTS) for this task. Using \textbf{LightGBM}, both \textbf{Pred-C} (0.5521) and \textbf{Pred-A} (0.6901) showed significantly lower Acc compared to the global contrastive approaches. This suggests that the predictive goal itself is not suitable for this task, regardless of how it predicts the future or past sequence. Since Seoul Korean tones are defined by the overall shape of the pitch, methods that rely on predicting only parts of the sequence fail to learn the global structure.

\paragraph{Efficacy of Global Constraints.}
All global contrastive models consistently exceed 0.76 Acc, outperforming predictive baselines by a wide margin. This confirms that applying \textbf{SupCon} on the full sequence effectively captures the holistic tonal patterns essential for classification.

\paragraph{Superiority of the Proposed Method.}
\textbf{Proposed model (Dual-Glob)} achieved the highest Acc (0.7775), showing that simply enforcing consistency between the global representations of different views is the most effective strategy. Interestingly, the \textbf{Hybrid} model (w/ LR) performed slightly worse (0.7712). This model combines the global contrastive objective with a predictive task that forecasts the augmented features from the clean ones. The results suggest that this explicit predictive constraint is unnecessary; it enforces the model to learn exact feature mappings between views, potentially distracting it from learning the invariant global patterns that are critical for classification.

Furthermore, our investigation into loss formulation strategies (Table~\ref{tab:ablation}) shows that the Dual-Glob approach yields favorable results compared to \textbf{Cross-View SupCon} (w/ LR) (0.7679) and \textbf{Unified SupCon} (w/ LR) (0.7732). While explicit cross-view alignment is common in self-supervised learning, these findings suggest that optimizing clean and augmented views through separate loss terms may provide a more balanced training signal for prosodic analysis. This independent supervision appears to encourage the model to capture the shared global patterns effectively while maintaining robustness against specific variations in perturbed pitch contours.

\begin{table}[t]
    \centering
    \caption{Performance comparison of contrastive models including loss formulation variants. Performance is measured via 5-fold cross-validation on frozen features.}
    \label{tab:ablation}
    \resizebox{\linewidth}{!}{
    \begin{tabular}{lcccccc}
        \toprule
        \multirow{2}{*}{\textbf{Method}} & \multicolumn{2}{c}{\textbf{LightGBM}} & \multicolumn{2}{c}{\textbf{RF}} & \multicolumn{2}{c}{\textbf{LR}} \\
        \cmidrule(lr){2-3} \cmidrule(lr){4-5} \cmidrule(lr){6-7}
        & \textbf{Acc} & \textbf{F1} & \textbf{Acc} & \textbf{F1} & \textbf{Acc} & \textbf{F1} \\
        \midrule
        \textbf{Pred-C} & 0.5521 & 0.3231 & 0.5469 & 0.3193 & 0.5275 & 0.32970 \\
        \textbf{Pred-A} & 0.6901 & 0.3722 & 0.5976 & 0.3549 & 0.5981 & 0.3064 \\
        \midrule
        \textbf{Glob-Clean}    & 0.7688 & 0.4892 & 0.7677 & 0.4822 & 0.7708 & 0.4931 \\
        \textbf{Glob-Augment}  & 0.7654 & 0.4838 & 0.7656 & 0.4844 & 0.7697 & 0.4918 \\
        \textbf{Hybrid}        & 0.7679 & 0.4956 & 0.7659 & 0.4868 & 0.7712 & 0.4947 \\
        \midrule
        \textbf{Cross-View SupCon}  & 0.7670 & 0.4877 & 0.7661 & 0.4835 & 0.7679 & 0.4878 \\
        \textbf{Unified SupCon}    & 0.7721 & 0.4970 & 0.7719 & 0.4950 & 0.7732 & 0.5051 \\
        \midrule
       
        \textbf{Proposed} (Dual-Glob)  & \textbf{0.7743} & \textbf{0.5051} & \textbf{0.7740} & \textbf{0.5051} & \textbf{0.7775} & \textbf{0.5154} \\
        \bottomrule
    \end{tabular}
    }
\end{table}

\begin{figure}[t]
    \centering
    \begin{subfigure}{0.48\linewidth}
        \centering
        \includegraphics[width=\linewidth]{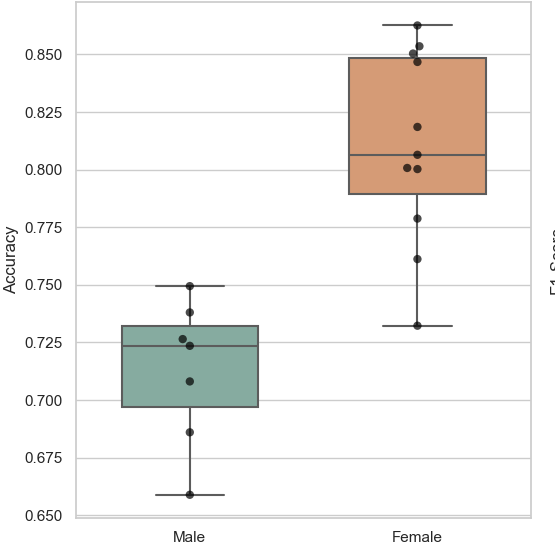} 
        \caption{Acc distribution.}
        \label{fig:gender_acc}
    \end{subfigure}
    \hfill
    \begin{subfigure}{0.48\linewidth}
        \centering
        \includegraphics[width=\linewidth]{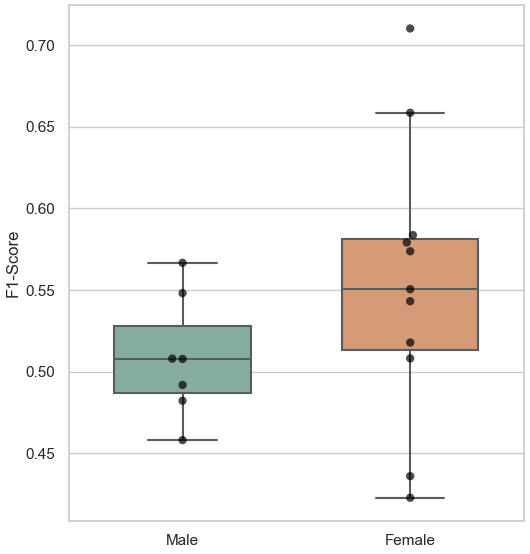} %
        \caption{F1 distribution.}
        \label{fig:gender_f1}
    \end{subfigure}
    
    \caption{Performance distribution of unified model across Acc and F1.}
    \vspace{-0.3cm}
    \label{fig:gender_performance}
\end{figure}

\paragraph{Effect of Gender Difference.}
Table \ref{tab:gender_analysis} shows that in the unified model, female speakers achieved significantly higher Acc than male speakers (\textbf{0.8075} vs. \textbf{0.7130}). As shown in Figure \ref{fig:gender_performance}, female speakers form a compact and high-performance cluster, whereas male speakers exhibit lower median Acc. Adopting a gender-specific approach further improved performance, increasing female speakers' Acc to \textbf{0.8120} and male speakers' Acc to \textbf{0.7288}. This difference is likely because female speakers use a wider pitch range, which makes their intonation patterns more distinct and easier for the model to recognize. In contrast, male speakers tend to use a narrower range, creating more similar and ambiguous patterns that are harder to tell apart. For a visual look at how these patterns are separated in the model, please see \textbf{Appendix E}.

\begin{table}[t]
    \centering
    \caption{Performance comparison between unified and gender-specific models.}
    \label{tab:gender_analysis}
    \resizebox{\linewidth}{!}{
    \begin{tabular}{lcccc}
        \toprule
        \multirow{2}{*}{\textbf{Gender}} & \multicolumn{2}{c}{\textbf{Unified Model}} & \multicolumn{2}{c}{\textbf{Gender-Specific Model}} \\
        \cmidrule(lr){2-3} \cmidrule(lr){4-5}
         & \textbf{Acc} & \textbf{F1} & \textbf{Acc} & \textbf{F1} \\
        \midrule
        \textbf{Male}   & 0.7130 \small{$\pm$0.04} & 0.4784 \small{$\pm$ 0.04} & 0.7288 \small{$\pm$ 0.02} & 0.5539 \small{$\pm$ 0.02} \\
        \textbf{Female} & \textbf{0.8075} \small{$\pm$ 0.01} & \textbf{0.5286} \small{$\pm$ 0.02} & \textbf{0.8120} \small{$\pm$ 0.02} & \textbf{0.6103} \small{$\pm$ 0.03} \\
        \bottomrule
    \end{tabular}
    }
\end{table}

\subsection{Ambiguity in Sustained Tones}

\begin{figure}[t]
    \centering
    \begin{subfigure}{1.0\linewidth}
        \centering
        \includegraphics[width=\linewidth]{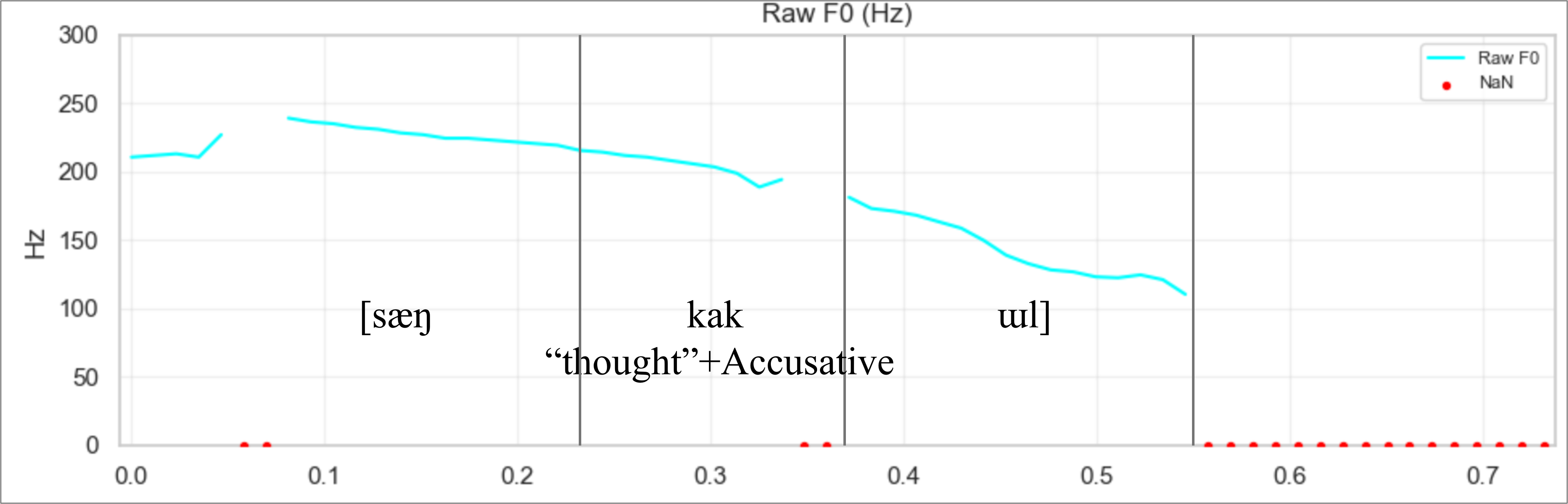}
        \caption{Misclassification of [\textipa{s\ae N.kak} + \textipa{\textturnm l}] ("thought"+Accusative): Predicted \text{HHLL}, Ground Truth \text{HHL}.}
        \label{fig:err_s}
    \end{subfigure}
    
    
    \begin{subfigure}{1.0\linewidth}
        \centering
        \includegraphics[width=\linewidth]{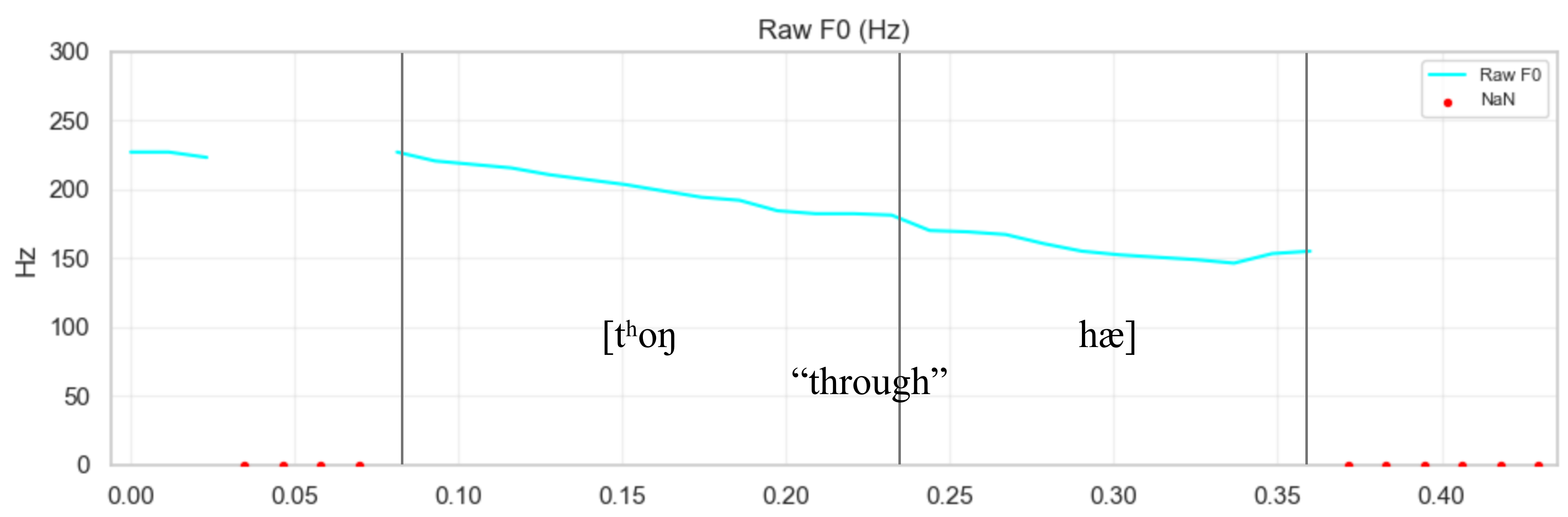}
        \caption{Misclassification of [\textipa{t\super{h}oN}.\textipa{h\ae}] ("through"): Predicted \text{HHLL}, Ground Truth \text{HL}.}
        \label{fig:err_t}
    \end{subfigure}
    
    \caption{Failure cases demonstrating the ambiguity in sustained tones. In both cases, the model misinterprets the lengthened final L tone as a sequence of multiple L tones (LL).}
    \label{fig:limitation_cases}
    \vspace{-0.3cm} 
\end{figure}

A detailed error analysis reveals a limitation in modeling tonal patterns ending with sustained L tones. Figure \ref{fig:limitation_cases} illustrates failure cases where the model incorrectly predicts \text{HHLL} for ground truth labels ending in single L tones.

In Figure \ref{fig:err_s}, the AP, [\textipa{s\ae N.kak} + \textipa{\textturnm l}] ("thought"+Accusative), labeled as \text{HHL}, is misclassified as \text{HHLL}. Similarly, in Figure \ref{fig:err_t}, the AP [\textipa{t\super{h}oN}.\textipa{h\ae}] ("through"), labeled as \text{HL}, is also predicted as \text{HHLL}. These errors come from the acoustic ambiguity of the final L tone. When a speaker lengthens the final syllable (e.g., [\textipa{h\ae}] in [\textipa{t\super{h}oNh\ae}]), the resulting long and flat $F_0$ contour is geometrically indistinguishable from a sequence of multiple L tones to a model that relies solely on $F_0$ shapes.

This ambiguity is because the model only looks at the $F_0$ shape and does not know the length of each sound or where the syllables end. Without this information, the model often mistakes a long and flat low pitch for several different patterns. For more details on these errors, please see \textbf{Appendix D}.

\section{Discussion}
\label{sec:discussion}

To address the aforementioned limitations, we incorporated syllable count constraints into the classification framework. Specifically, we encoded the syllable count ($N_{syl}$) of each AP into a structural vector. We clip the maximum length at 4 (i.e., any $N_{syl} \ge 4$ is encoded as the maximum category).

This syllable information is encoded as a one-hot vector $v_{syl} \in \mathbb{R}^{4}$ and concatenated with the frozen latent representation $z$ derived from the encoder. The final input to the classifier becomes $[z; v_{syl}]$. This fusion strategy allows the model to distinguish whether a sustained contour corresponds to a single lengthened syllable or multiple tonal targets, thereby resolving the ambiguity in patterns like \text{HL} versus \text{HHLL}.

The results in Table \ref{tab:model_results} show the performance of our syllable-aware model. By combining syllable counts with $F_0$, the model reached its highest Acc of 0.894 with LR. This demonstrates that adding simple timing information helps the model better understand complex intonation patterns. 

We also analyzed how the model performs for different genders in Table \ref{tab:gender_analysis2}. The syllable-aware model with LightGBM worked particularly well for female speakers, achieving 0.9154 Acc and 0.7725 F1. This confirms that even after adding syllable information, the clearer and wider pitch ranges of female voices continue to provide more distinct cues for the model compared to male voices.

\begin{table}[t]
\centering
\small 
\caption{Classification performance comparison of the proposed syllable-aware model.}
\label{tab:model_results}
\begin{tabular}{lcc}
\toprule
\textbf{Ours} & \textbf{Acc} & \textbf{F1} \\
\midrule
w/ LightGBM      & 0.891 $\pm$ 0.014 & 0.694 $\pm$ 0.020 \\
w/ RF & 0.865 $\pm$ 0.015          & 0.622 $\pm$ 0.024 \\
w/ LR & \textbf{0.894 $\pm$ 0.021}          & \textbf{0.689 $\pm$ 0.013} \\
\bottomrule
\end{tabular}
\end{table}

\begin{table}[t]
\centering
\caption{Gender-specific performance analysis for the proposed syllable-aware model with LightGBM.}
\label{tab:gender_analysis2}
\begin{tabular}{lcc}
\toprule
\multirow{2}{*}{Gender} & \multicolumn{2}{c}{Ours} \\
\cmidrule(r){2-3}
 & Acc & F1 \\
\midrule
\textbf{Male}   & 0.8548 $\pm$ 0.04 & 0.7153 $\pm$ 0.02 \\
\textbf{Female} & 0.9154 $\pm$ 0.02 & 0.7725 $\pm$ 0.03 \\
\bottomrule
\end{tabular}
\end{table}

\section{Limitations}
Although the proposed model effectively classifies pitch contours, several factors limit the extent to which tonal categories can be fully captured. First, while we rely on $F_0$, intonation is closely related to other prosodic cues, such as segmental duration and intensity that also mark prominence.

In addition, $F_0$ measurement itself remains imperfect: (1) $F_0$ track is often lost due to vowel devoicing or at the phrase-final position, (2) pitch halving errors (sudden drop of $F_0$) may lead to tonal misinterpretation, (3) glottalization in phrase-final L tones lead to pitch tracking loss or errors, and (4) $F_0$ rise due to local pitch perturbations following the release of the obstruents. (For a detailed discussion on these measurement difficulties, see \textbf{Appendix F}.)

Furthermore, as illustrated in Table 1, the distribution of pitch labels exhibits a significant class imbalance. In particular, categories with fewer than 100 samples show relatively low precision and recall. Consequently, overall F1 across all models, including the baselines, remain constrained. To address this, it is essential to secure a more substantial dataset, for example, obtaining several hundred samples per class, and to employ specialized loss functions or training strategies designed for class-balancing. For a more detailed error analysis for each class, please refer to Appendix D.(b).


Despite limitations, the model performs strongly using $F_0$ alone, suggesting pitch contours encode substantial structural information. Note that our analysis highlights inherent gender differences: female speakers typically exhibit wider ranges and more dynamic excursions in $F_0$ than male speakers.

However, it should be noted our dataset was constructed based on broadcast speech, which is optimized for clear information delivery, but this may lead to a restricted set of pitch patterns, potentially constraining the observed variability of pitch patterns. Given that \citet{kim2008intonational} found that pitch accent distributions differ by register, future work should therefore examine a broader range of speech styles and corpora to validate the generalizability of these findings.

\section{Conclusion}
In this work, we make two main contributions to the study of Seoul Korean intonation. First, we constructed a \textbf{new benchmark dataset} by manually labeling AP in Seoul Korean. This dataset provides high-quality tonal labels, making it a rich resource for various fields such as speech synthesis and automated linguistic analysis.

Second, we presented \textbf{Dual-Glob}, a framework designed to capture the holistic shape of $F_0$ contours. Unlike previous methods that focus on local transitions, our approach learns the entire structure of the pitch pattern at once. Our results showed that this global perspective is much more effective, achieving significantly higher Acc than existing models.

Our analysis also revealed that while the narrower pitch ranges of male speakers can make classification more difficult, our model remains robust by capturing the global structure of the sounds. These findings confirm that focusing on the $F_0$ shapes as a whole is the most reliable way to analyze intonational patterns.

\clearpage      
\bibliography{custom}

\appendix
\clearpage      
\onecolumn      
\appendix       
\section{Appendix A: Schematic $F_0$ contours of an AP}
\label{sec:appendix_A}

\begin{figure}[h]
    \centering
    \includegraphics[width=1.0\linewidth]{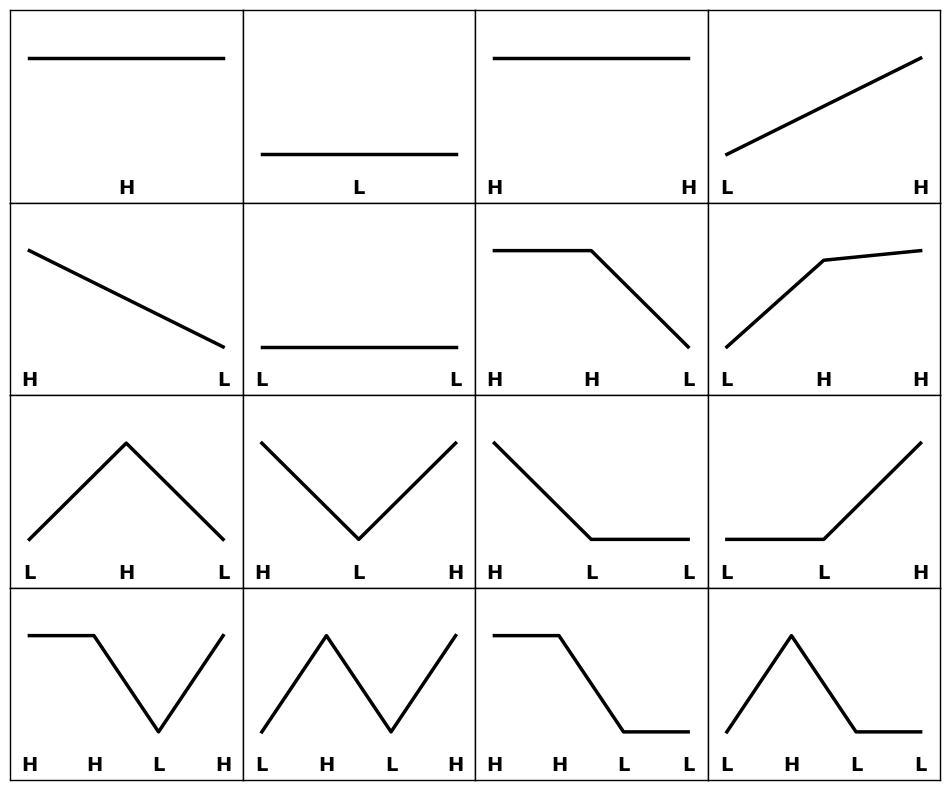}
    \caption{Schematic $F_0$ contours of sixteen pitch accent patterns for an AP in Seoul Korean \cite{jun2000k} }
    \label{fig:schematic_patterns}
\end{figure}

The schematic $F_0$ contours of an AP in Seoul Korean are provided in Figure \ref{fig:schematic_patterns} (Redrawn from \citet{jun2000k}). The APs in Seoul Korean are typically realized with the following sixteen pitch accent categories.

\section*{Appendix B: Implementation Details}

\begin{table*}[ht!]
    \centering
    \caption{Detailed architecture configurations and hyperparameters for all implemented models. All baselines were trained end-to-end, whereas the proposed method employed a fixed feature extractor followed by separate classifiers.}
    \label{tab:model_architectures}
    \resizebox{0.95\textwidth}{!}{
    \begin{tabular}{l|l|l}
        \toprule
        \textbf{Model} & \textbf{Component} & \textbf{Configuration Details} \\
        \midrule
        \multirow{4}{*}{\textbf{Proposed (Dual-Glob)}} & Encoder & 6-layer Conv1D (Kernels=[16, 12, 9, 6, 6, 6], Channels=[16, 32, 64, 128, 256, $D_{emb}$], Strides=[1, 2, 2, 1, 1, 1]) + Masked GAP  \\
         & Projector & 2-layer MLP (Hidden=64, Output=64, ReLU) \\
         & Predictor & 2-layer MLP (Hidden=64, Output=64, ReLU) \\
         & Classifier & Logistic Regression, Random Forest ($n_{est}=200$), LightGBM ($n_{est}=200$) \\
        \midrule
      \multirow{2}{*}{\textbf{CNN}} & Backbone & 5 Conv1D blocks (Channels=[64, 128, 256, 1024], Kernel=[16, 12, 9, 9]) + MaxPool(2) \\
         & Head & AdaptiveAvgPool1d(1) $\to$ Linear(1024 $\to$ $N_{classes}$) \\
        \midrule
        \multirow{2}{*}{\textbf{LSTM}} & Backbone & Bidirectional LSTM (Hidden=64, Layers=2, BatchFirst=True) \\
         & Head & Concatenated last hidden states (128 dim) $\to$ Linear(128 $\to$ $N_{classes}$) \\
        \midrule
        \multirow{2}{*}{\textbf{InceptionTime}} & Backbone & 6 Inception Blocks (Depth=6, Hidden=128) + Residual Connections \\
         & Head & AdaptiveAvgPool1d(1) $\to$ Flatten $\to$ Linear(512 $\to$ $N_{classes}$) \\
        \midrule
        \multirow{2}{*}{\textbf{Transformer}} & Backbone & 2 Encoder Layers ($d_{model}=256, n_{head}=4, d_{ff}=256$) + Positional Encoding \\
         & Head & Global Average Pooling $\to$ Linear(256 $\to$ $N_{classes}$) \\
        \midrule
        \multirow{2}{*}{\textbf{MiniRocket}} & Transform & 50,000 random convolutional kernels (Fixed weights) \\
         & Classifier & Ridge Classifier (Linear, $\alpha \in [10^{-3}, 10^{3}]$) \\
        \midrule
        \textbf{TimesNet} & Architecture & 2 Layers, $d_{model}=32$, Top-$k$=1, Multi-periodicity analysis \\
        \midrule
        \textbf{DLinear} & Architecture & Single Linear Layer decomposition (SeqLen=200) \\
        \bottomrule
    \end{tabular}
    }
\end{table*}
\paragraph{Environment and Data Split.}
All models were implemented using \textbf{PyTorch} and trained on an NVIDIA GPU RTX 2070. To ensure a robust evaluation, we employed \textbf{5-fold stratified cross-validation} with a fixed random seed (42). For each fold, the dataset was split into training and validation sets, and the final performance is reported as the mean and standard deviation across the five folds.

\paragraph{Model Architecture.}
For the proposed \textbf{Dual-Glob} framework, we utilized a \textbf{6-layer 1D CNN encoder} with kernel sizes of $[16, 12, 9, 6, 6, 6]$, strides of $[1, 2, 2, 1, 1, 1]$, and increasing channel sizes of $[16, 32, 64, 128, 256, D_{emb}]$. After the convolutional layers, a masked global average pooling (GAP) was applied to obtain the latent representation. Both the \textbf{projection and prediction heads} were configured as 2-layer MLPs with a hidden and output dimension of 64 and ReLU activation. To evaluate the robustness of the learned embeddings, we systematically varied the \textbf{embedding size ($D_{emb}$)} from \textbf{64 to 1024} during our experiments. For the total objective function, the weighting hyperparameters $\lambda_1$ and $\lambda_2$ were both set to 1. 

To enhance the density of positive and negative pairs within each training iteration, we adopted a batch-duplication strategy where each mini-batch was concatenated with itself. This approach effectively increases the number of contrastive relations, contributing to more stable and robust gradient estimation without requiring additional unique data samples.

For baseline comparisons, we deployed standard implementations of CNN, LSTM, InceptionTime, Transformer, DLinear, and TimesNet, along with MiniRocket. Deep learning baselines were trained end-to-end, whereas MiniRocket was evaluated using a Ridge Classifier.

\paragraph{Training Configuration.}
We optimized the models using \textbf{RAdam} combined with the \textbf{Lookahead} mechanism ($k=5, \alpha=0.9$) to stabilize convergence. The learning rate was set to $1 \times 10^{-2}$ for self-supervised pre-training and $3 \times 10^{-3}$ for supervised baselines, with a weight decay of $1 \times 10^{-4}$. The batch size was fixed at 64.

The training duration varied by model, ranging from 50 to 100 epochs depending on convergence speed. To ensure reliable performance estimation and account for training fluctuations, we calculated the final metrics by averaging the results of the last 5 training epochs across all 5 folds, rather than relying on a single best-epoch checkpoint.

Table \ref{tab:model_architectures} summarizes the specific architecture configurations and hyperparameter settings for all implemented models.

\section*{Appendix C: Effect of Data Augmentation}

For contrastive learning, we employed a stochastic data augmentation strategy to learn noise-invariant representations. For each training instance, we \textbf{randomly selected 3 transformations} from the following pool of five techniques:
\begin{itemize}
    \item \textbf{Random Jittering (J):} Adding Gaussian noise ($\sigma=0.02$) to simulate recording noise.
    \item \textbf{Scaling (S):} Multiplying the amplitude by a random factor ($0.8 \sim 1.2$) to handle intensity variations.
    \item \textbf{Masking (M):} Randomly zeroing out a portion of the sequence (ratio=0.2) to encourage robustness against missing data.
    \item \textbf{Magnitude Shift:} Adding a random constant bias to the Log-$F_0$ contour, simulating global pitch level shifts.
    \item \textbf{Time Warping:} Applying non-linear temporal deformation using random knots and linear interpolation to simulate local speaking rate variations.
\end{itemize}
To evaluate the impact of different augmentation selection strategies on model performance, we conducted comparative experiments across six distinct configurations (D1--D6). These configurations vary in the number of transformations selected and the diversity of the augmentation pool.

As shown in Table \ref{tab:comparative_results} the D4 strategy (2$\sim$3 random selection) consistently achieves the highest overall performance across all classifiers. It reaches a peak of 77.75\% Acc and 51.54\% F1 with the LR model and maintains the highest average performance (77.53\% Acc / 50.97\% F1). These results demonstrate that the dynamic combination of 2 to 3 augmentations is the most effective approach for learning robust and discriminative pitch accent representations.

\begin{table}[ht]
\centering
\caption{Comparative results of 5-fold cross validation across different adapter selection strategies (Encoder Dimension: 1024). Acc and F1 are reported in percentages (\%).}
\label{tab:comparative_results}
\resizebox{\columnwidth}{!}{%
\begin{tabular}{@{}llcccc@{}}
\toprule
\textbf{Dataset} & \textbf{Selection Strategy} & \textbf{RF (Acc/F1)} & \textbf{LR (Acc/F1)} & \textbf{LGBM (Acc/F1)} & \textbf{Avg. (Acc/F1)} \\ \midrule
D1 & 1 random (Full) & 77.25 / 50.12 & 77.49 / 51.25 & 77.48 / 50.39 & 77.41 / 50.59 \\
D2 & 2 random (Full) & 77.39 / 49.96 & 77.64 / 51.25 & 77.30 / 50.21 & 77.44 / 50.47 \\
D3 & 3 random (Full) & 77.23 / 49.74 & 77.61 / 50.35 & 77.24 / 50.78 & 77.36 / 50.29 \\
\textbf{D4 (Ours)} & \textbf{2$\sim$3 random (Full)} & \textbf{77.40 / 50.51} & \textbf{77.75 / 51.54} & \textbf{77.43 / 50.86} & \textbf{77.53 / 50.97} \\
D5 & 1 random \{J, S, M\} & 77.09 / 49.53 & 77.47 / 50.87 & 77.23 / 50.04 & 77.26 / 50.15 \\
D6 & 2 random \{J, S, M\} & 77.30 / 49.99 & 77.57 / 50.81 & 77.60 / 51.18 & 77.48 / 50.68 \\ \bottomrule
\end{tabular}%
}
\end{table}

To further evaluate the effectiveness of the proposed approach, we investigated the impact of data augmentation (D4) on various baseline models. For these baseline experiments, we trained the models by concatenating the original clean data with the augmented data, effectively doubling the training size to ensure a fair comparison. As summarized in Table \ref{tab:augmentation_results}, the application of data augmentation generally led to performance gains across most architectures. Specifically, models such as BiLSTM, InceptionTime, and Transformer exhibited notable increases in both Acc and F1. These results suggest that our data augmentation strategy contributes to better generalization by allowing models to learn from a more diverse set of pitch contour variations, even when applied to conventional supervised learning frameworks.

\begin{table}[htbp]
\centering
\caption{Performance comparison of baseline models with (w/) and without (w/o) data augmentation. The w/ Augmentation setting uses a concatenation of clean and augmented data. Results represent the average of 5-fold cross-validation (Mean $\pm$ SD).}
\label{tab:augmentation_results}
\resizebox{\columnwidth}{!}{%
\begin{tabular}{lcccc}
\toprule
\multirow{2}{*}{\textbf{Model}} & \multicolumn{2}{c}{\textbf{w/o Augmentation}} & \multicolumn{2}{c}{\textbf{w/ Augmentation}} \\
\cmidrule(lr){2-3} \cmidrule(lr){4-5}
& \textbf{Acc} & \textbf{F1} & \textbf{Acc} & \textbf{F1} \\
\midrule
1D-CNN         & 0.7291 $\pm$ 0.0087 & \textbf{0.5011} $\pm$ 0.0157 & \textbf{0.7410} $\pm$ 0.0104 & 0.4930 $\pm$ 0.0134 \\
BiLSTM         & 0.7440 $\pm$ 0.0087 & 0.4782 $\pm$ 0.0190          & \textbf{0.7568} $\pm$ 0.0156 & \textbf{0.4915} $\pm$ 0.0186 \\
InceptionTime  & 0.7327 $\pm$ 0.0071 & 0.4915 $\pm$ 0.0196          & \textbf{0.7426} $\pm$ 0.0106 & \textbf{0.5043} $\pm$ 0.0147 \\
DLinear        & 0.6237 $\pm$ 0.0070 & 0.3875 $\pm$ 0.0139          & \textbf{0.6461} $\pm$ 0.0078 & \textbf{0.3892} $\pm$ 0.0242 \\
MiniRocket     & 0.7287 $\pm$ 0.0159 & \textbf{0.4372} $\pm$ 0.0236 & \textbf{0.7303} $\pm$ 0.0152 & 0.4322 $\pm$ 0.0179 \\
Transformer    & 0.6943 $\pm$ 0.0770 & 0.4434 $\pm$ 0.0133          & \textbf{0.7177} $\pm$ 0.0107 & \textbf{0.4680} $\pm$ 0.0248 \\
\bottomrule
\end{tabular}%
}
\end{table}

\newpage

\section*{Appendix D: Detailed Error Analysis}
\label{sec:appendix_b}

\begin{figure}[h]
    \centering
    \includegraphics[width=0.7\linewidth]{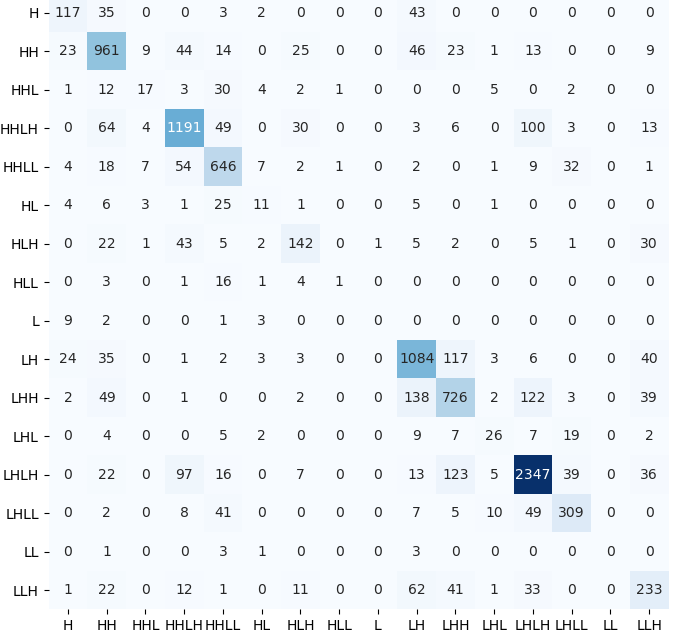}
    \caption{Confusion matrix of the proposed Dual-Glob method with LR. The strong diagonal density confirms robust classification accuracy across all tonal classes.}
    \vspace{-0.2cm} 
    \label{fig:confusion_matrix}
\end{figure}

To quantitatively evaluate the classification behavior, we analyze the confusion matrix shown in Figure \ref{fig:confusion_matrix}. The high concentration of values along the diagonal indicates that the model successfully distinguishes distinct tonal classes in most cases. However, off-diagonal elements reveal that misclassifications are not random but primarily occur between similar patterns (e.g., \text{HL} vs. \text{HHLL}). These confusions suggest that while the global $F_0$ contour shapes are well-captured, subtle ambiguities in duration and syllable boundaries remain challenging.

\subsection*{(a) Analysis of Error Patterns}

We qualitatively analyzed failure cases where the model misclassifies tonal patterns. In each figure below, the upper and lower panels show representative examples of the error patterns. These error patterns are largely coming from the ambiguous situations when interpreting sustained tonal contours or lengthened syllables without segmenting syllable boundaries.

\begin{figure}[t]
    \centering
    \begin{subfigure}{0.48\linewidth}
        \centering
        \includegraphics[width=\linewidth]{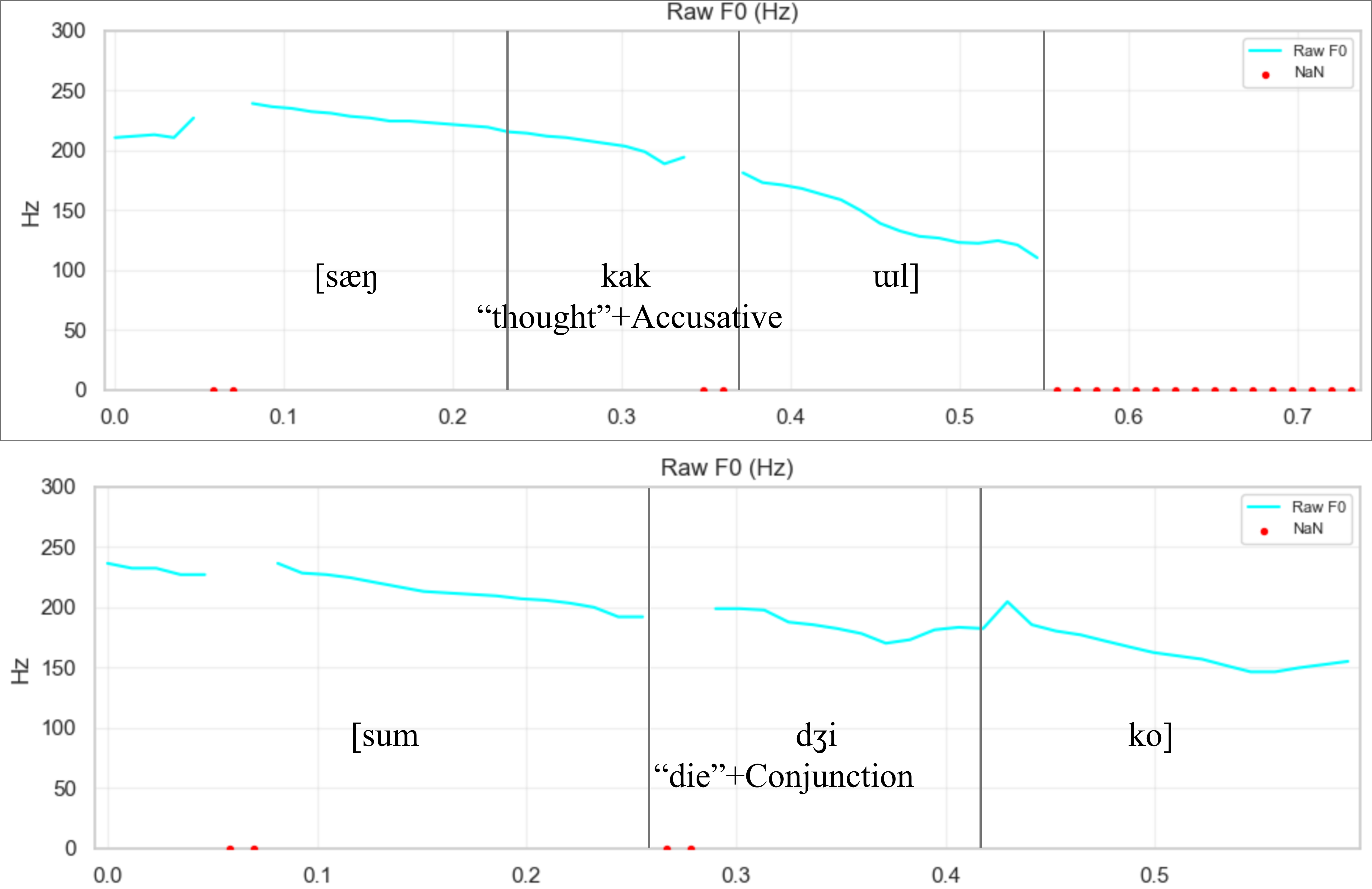}
        \caption{\text{HHL} $\rightarrow$ \text{HHLL}. \\ 
        Top: [\textipa{s\ae N.kak} + \textipa{\textturnm l}] ("thought"+Accusative) \\ 
        Bottom: [\textipa{sum.d\textyogh i.ko}]  ("die"+Conjunction)}
        \label{fig:err0}
    \end{subfigure}
    \hfill
    \begin{subfigure}{0.48\linewidth}
        \centering
        \includegraphics[width=\linewidth]{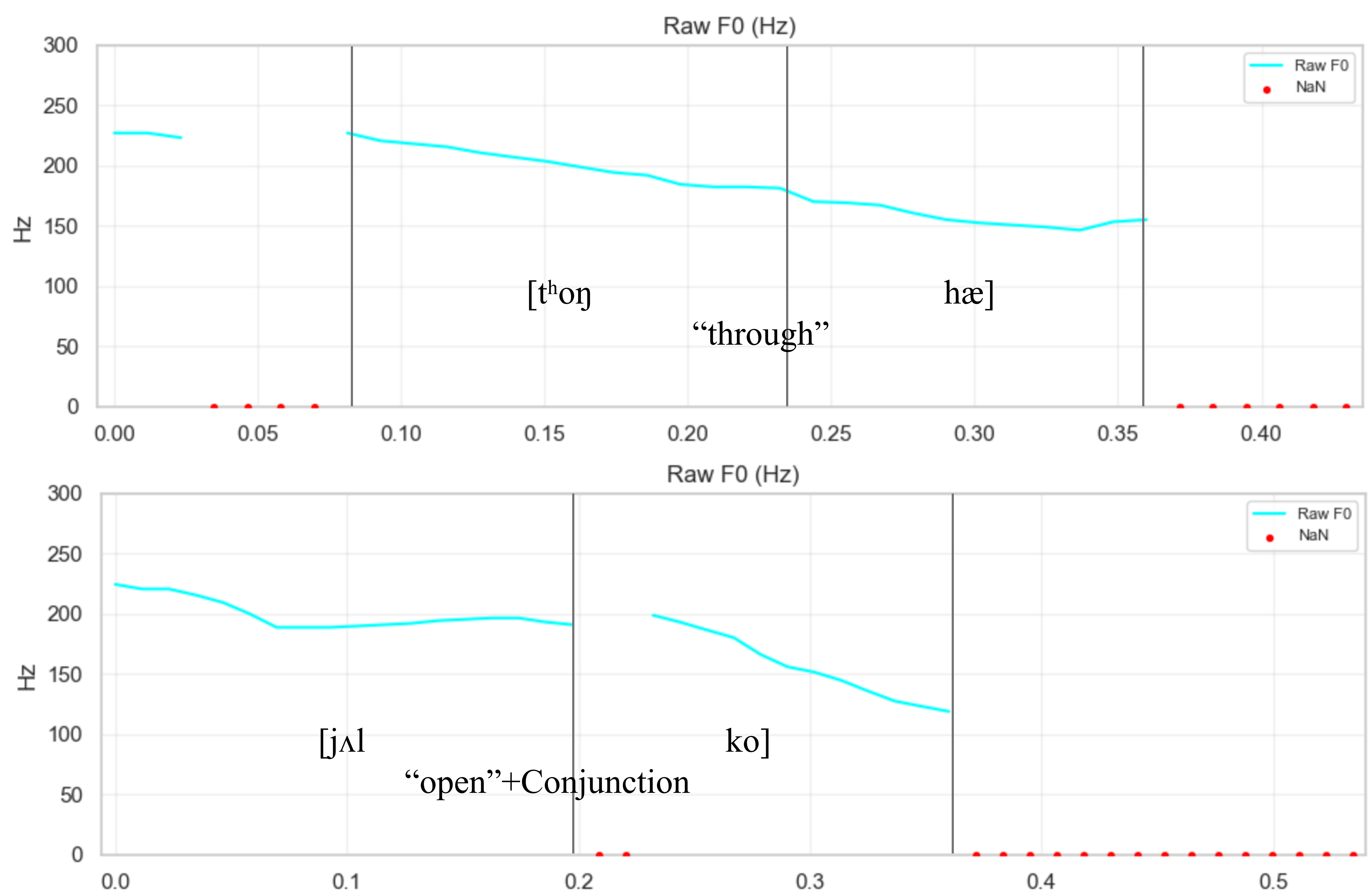}
        \caption{\text{HL} $\rightarrow$ \text{HHLL}. \\ 
        Top: [\textipa{t\super{h}oN}.\textipa{h\ae}] ("through") \\ 
        Bottom: [\textipa{j\textturnv l.ko}] ("open"+Conjunction)}
        \label{fig:err1}
    \end{subfigure}
    
    \vspace{0.5cm}
    
    \begin{subfigure}{0.48\linewidth}
        \centering
        \includegraphics[width=\linewidth]{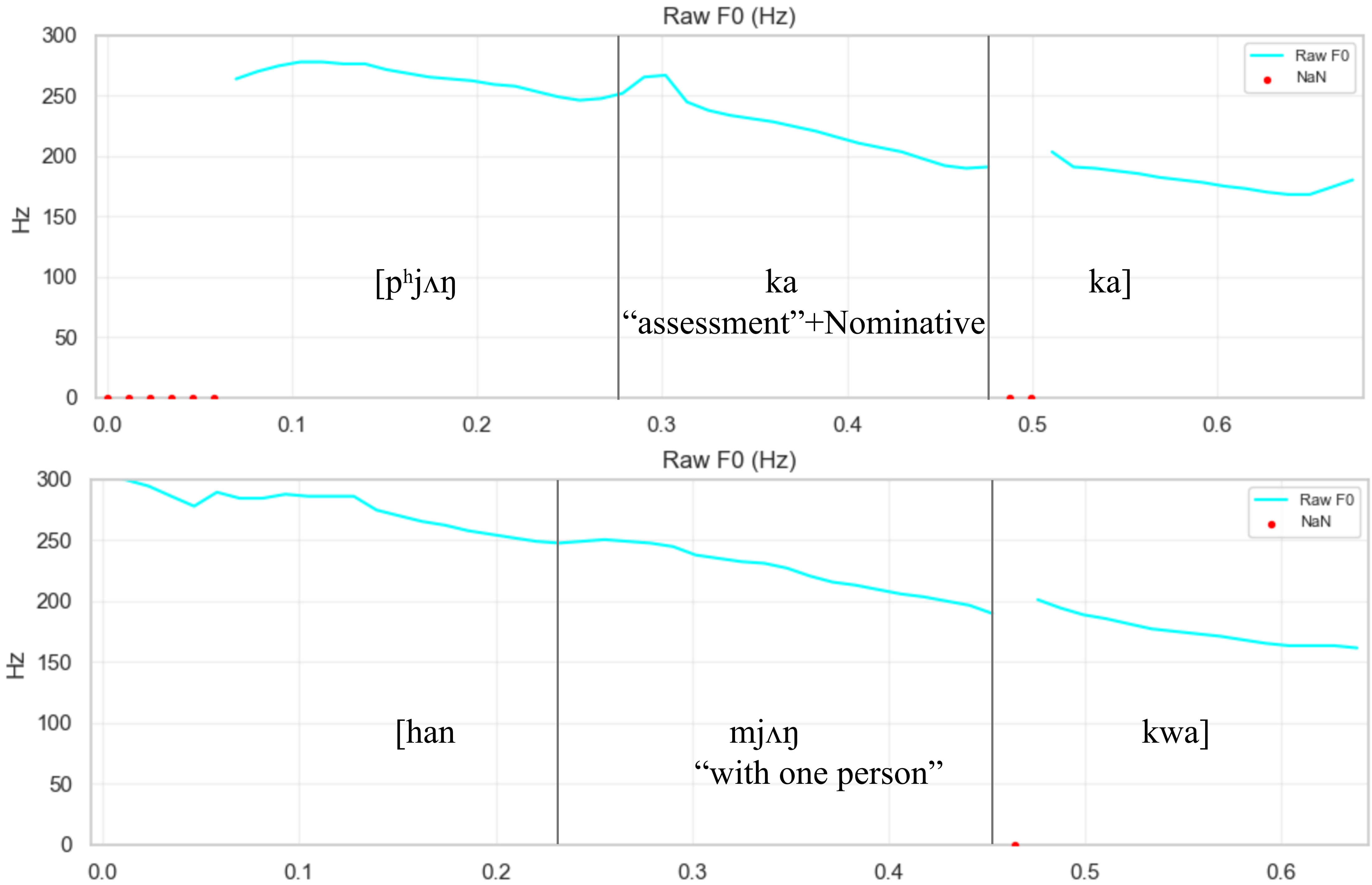}
        \caption{\text{HLL} $\rightarrow$ \text{HHLL}. \\ 
        Top: [\textipa{p\super{h}j\textturnv N.ka.ka}] ("assessment"+Nominative) \\ 
        Bottom: [\textipa{han.mj\textturnv N.kwa}] ("with one person")}
        \label{fig:err2}
    \end{subfigure}
    \hfill
    \begin{subfigure}{0.48\linewidth}
        \centering
        \includegraphics[width=\linewidth]{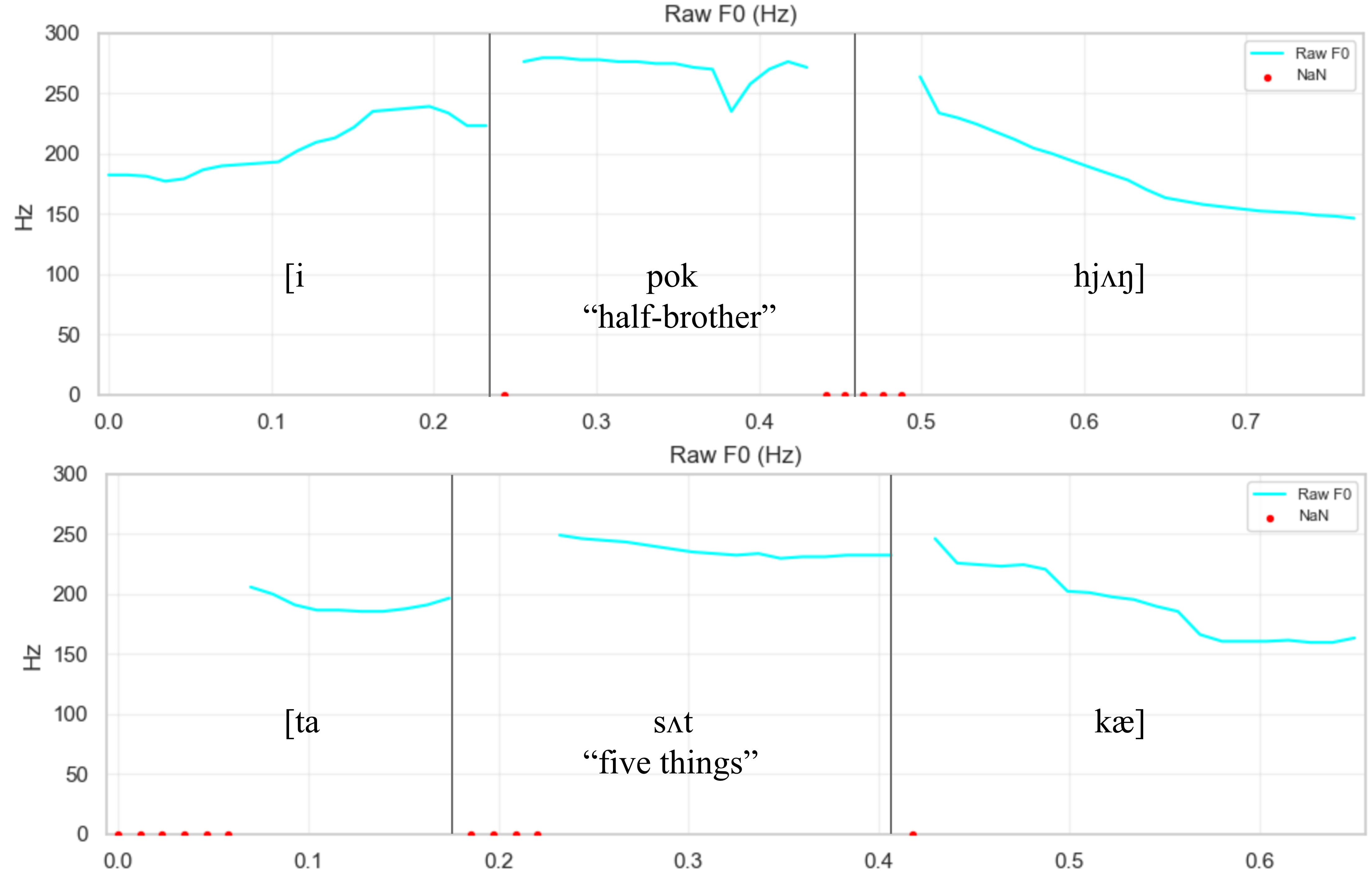}
        \caption{\text{LHL} $\rightarrow$ \text{LHLL}. \\ 
        Top: [\textipa{i.pok.hj\textturnv N}] ("half-brother") \\ 
        Bottom: [\textipa{ta.s\textturnv t.k\ae}] ("five things")}
        \label{fig:err3}
    \end{subfigure}
    
    \caption{Visualization of common misclassification patterns. Each subfigure displays two examples (Top/Bottom) showing the same type of prediction error. The vertical lines indicate the ground-truth syllable boundaries, which are hidden from the model.}
    \label{fig:appendix_errors}
\end{figure}

\begin{sloppypar}
\textbf{(1) \text{HHL} $\rightarrow$ \text{HHLL} (Figure \ref{fig:err0}):}
This category shows APs with three syllables, which is misclassified as four tones. In the example on the top, [\textipa{s\ae N.kak} + \textipa{\textturnm l}] ("thought"+Accusative), the final syllable [\textipa{\textturnm l}] is lengthened. Similarly, in the example at the bottom, [\textipa{sum.d\textyogh i.ko}]  ("die"+Conjunction), the final syllable [ko] shows a sustained contour. Due to these reasons, the model incorrectly splits these final segments into two low tones (\text{LL}).

\textbf{(2) \text{HL} $\rightarrow$ \text{HHLL} (Figure \ref{fig:err1}):}
These are APs with two syllables, but misclassifed as four tones. In [\textipa{t\super{h}oN}.\textipa{h\ae}] ("through") (Top), the final [\textipa{h\ae}] is lengthened. In [\textipa{j\textturnv l.ko}] ("open"+Conjunction)(Bottom), the initial syllable [\textipa{j\textturnv l}] is also long, which results in a lengthy flat trajectory. The model misinterprets these longer durations as multiple tonal targets, predicting \text{HHLL} for both.

\textbf{(3) \text{HLL} $\rightarrow$ \text{HHLL} (Figure \ref{fig:err2}):}
Both [\textipa{p\super{h}j\textturnv N.ka.ka}] ("assessment"+Nominative) (Top) and [\textipa{han.mj\textturnv N.kwa}] ("with one person") are APs with three syllables ending in two L tones. The continuous low-pitch part for the last two syllables (e.g., [ka.ka]' or `[\textipa{mj\textturnv N.kwa}]) does not show a clear acoustic dip. Without syllable boundaries, the model fails to match the $F_0$ contour with each L tone, leading to an \text{HHLL} pattern.

\textbf{(4) \text{LHL} $\rightarrow$ \text{LHLL} (Figure \ref{fig:err3}):}
This error occurs in \text{LHL} patterns where the final Low is misrecognized. In [\textipa{i.pok.hj\textturnv N}] ("half-brother") (Top) and [\textipa{ta.s\textturnv t.k\ae}] ("five things") (Bottom), the final syllables ([\textipa{hj\textturnv N}]  and \textipa{k\ae}]) exhibit a falling or sustained L tone. The model correctly identifies the initial \text{LH} rise but misinterpret the final L into \text{LL}, resulting in an \text{LHLL} prediction.
\end{sloppypar}

\subsection*{(b) Per-Class Metrics and Error Analysis}

Table \ref{tab:class_report} provides a detailed breakdown of the classification performance for each pitch accent class. While the macro-averaged F1 of the proposed model is approximately 0.51, which may appear relatively modest, a closer inspection of the per-class metrics reveals that this is primarily driven by the extreme data sparsity in minority categories.

As illustrated in Table \ref{tab:class_report}, classes with a small number of samples, such as HHL, HL, HLL, LHL, L, and LL show significantly lower performance, which heavily penalizes the macro-average. In contrast, dominant labels with over 1,000 samples, including HH, HHLH, LH, LHH, and LHLH, consistently achieve F1-scores ranging from 0.68 to as high as 0.87. These results indicate that the model achieves robust performance on dominant prosodic patterns, while the lower macro-average is primarily a reflection of the inherent data imbalance.

For future research, we aim to construct a more extensive and balanced dataset to mitigate these issues. Additionally, we plan to incorporate advanced training strategies, such as class-balanced loss or cost-sensitive learning, to further improve the model's sensitivity to uncommon prosodic patterns.

\begin{table}[t]
\centering
\small
\caption{Detailed classification performance for each pitch accent class in Seoul Korean. The results are evaluated via 5-fold cross-validation.}
\label{tab:class_report}
\begin{tabular}{lcccc}
\toprule
\textbf{Tone Class} & \textbf{Precision} & \textbf{Recall} & \textbf{F1} & \textbf{Support} \\ \midrule
H                   & 0.63               & 0.58            & 0.61              & 200              \\
HH                  & 0.76               & 0.82            & 0.79              & 1,168             \\
HHL                 & 0.41               & 0.22            & 0.29              & 77               \\
HHLH                & 0.82               & 0.81            & 0.82              & 1,463             \\
HHLL                & 0.75               & 0.82            & 0.79              & 784              \\
HL                  & 0.31               & 0.19            & 0.24              & 57               \\
HLH                 & 0.62               & 0.55            & 0.58              & 259              \\
HLL                 & 0.33               & 0.04            & 0.07              & 26               \\
L                   & 0.00               & 0.00            & 0.00              & 15               \\
LH                  & 0.76               & 0.82            & 0.79              & 1,318             \\
LHH                 & 0.69               & 0.67            & 0.68              & 1,084             \\
LHL                 & 0.47               & 0.32            & 0.38              & 81               \\
LHLH                & 0.87               & 0.87            & 0.87              & 2,705             \\
LHLL                & 0.76               & 0.72            & 0.74              & 431              \\
LL                  & 0.00               & 0.00            & 0.00              & 8                \\
LLH                 & 0.58               & 0.56            & 0.57              & 417              \\ \midrule
\textbf{Accuracy}   & \multicolumn{3}{c}{\textbf{0.77}}                      & 10,093           \\
Macro Avg.          & 0.55               & 0.50            & 0.51              & 10,093           \\
Weighted Avg.       & 0.77               & 0.77            & 0.77              & 10,093           \\ \bottomrule
\end{tabular}
\end{table}

\newpage
\section*{Appendix E: Latent Space Visualization by Gender}
\label{sec:appendix_gender_tsne}
\begin{figure}[ht]
    \centering
    \begin{subfigure}[b]{0.46\linewidth} 
        \centering
        \includegraphics[width=\linewidth]{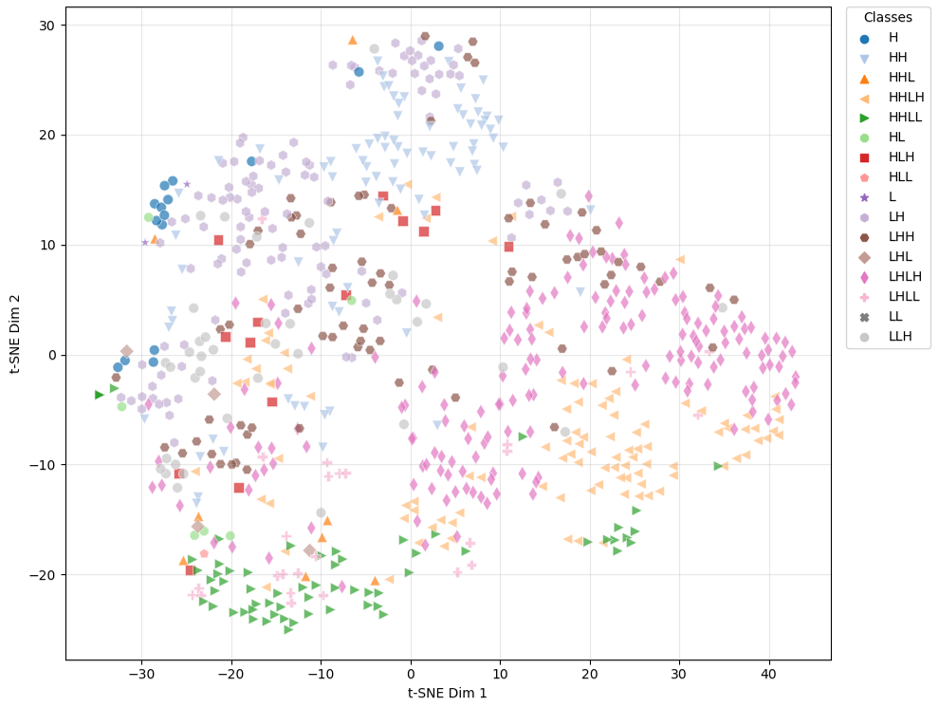}
        \caption{Male Speakers}
        \label{fig:tsne_male}
    \end{subfigure}
    \hfill 
    \begin{subfigure}[b]{0.46\linewidth} 
        \centering
        \includegraphics[width=\linewidth]{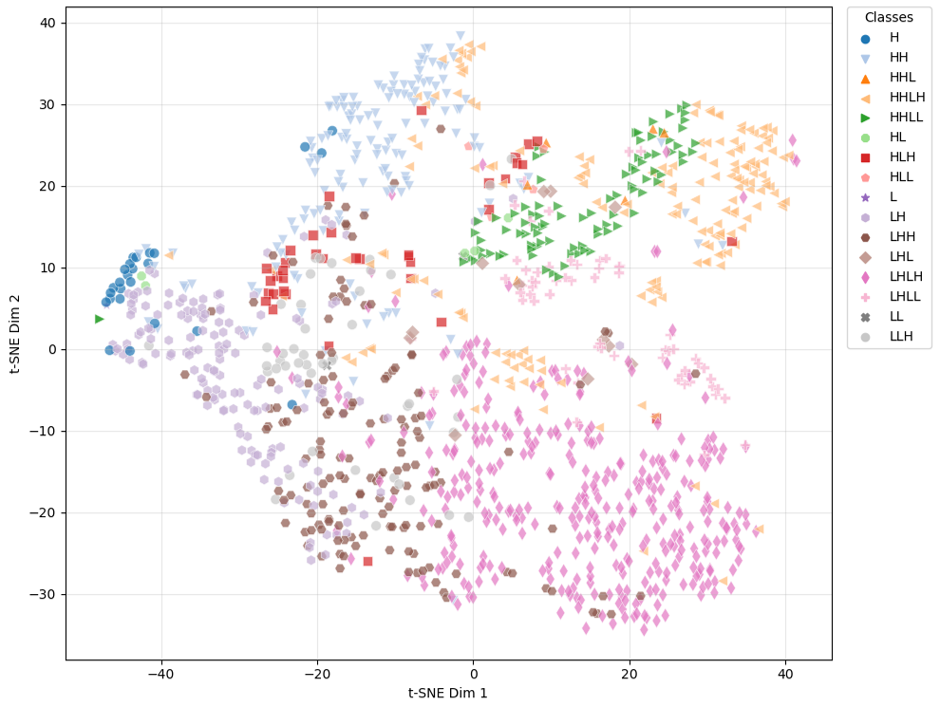}
        \caption{Female Speakers}
        \label{fig:tsne_female}
    \end{subfigure}
    
    \vspace{-0.2cm}
    
    \caption{t-SNE visualization of the feature space learned by the proposed model. Female speakers (b) show distinct class separation, whereas male speakers (a) exhibit significant overlap and scattering.}
    \label{fig:gender_tsne}
    
    \vspace{-0.5cm} 
\end{figure}

We visualized learned representations via t-SNE (Figure \ref{fig:gender_tsne}). Figure \ref{fig:gender_tsne}b shows the female model forms well-separated clusters, indicating wider pitch ranges provide clear cues. In contrast, Figure \ref{fig:gender_tsne}a reveals that the male space overlaps due to narrower pitch excursions, creating ambiguous boundaries between similar contours.


\section*{Appendix F: Difficulties in Pitch Tracking}
\label{sec:appendix_pitch_tracking}

\begin{figure}[p]
    \centering
    
    \begin{subfigure}[b]{0.48\linewidth}
        \centering
        \includegraphics[width=\textwidth, height=4.5cm]{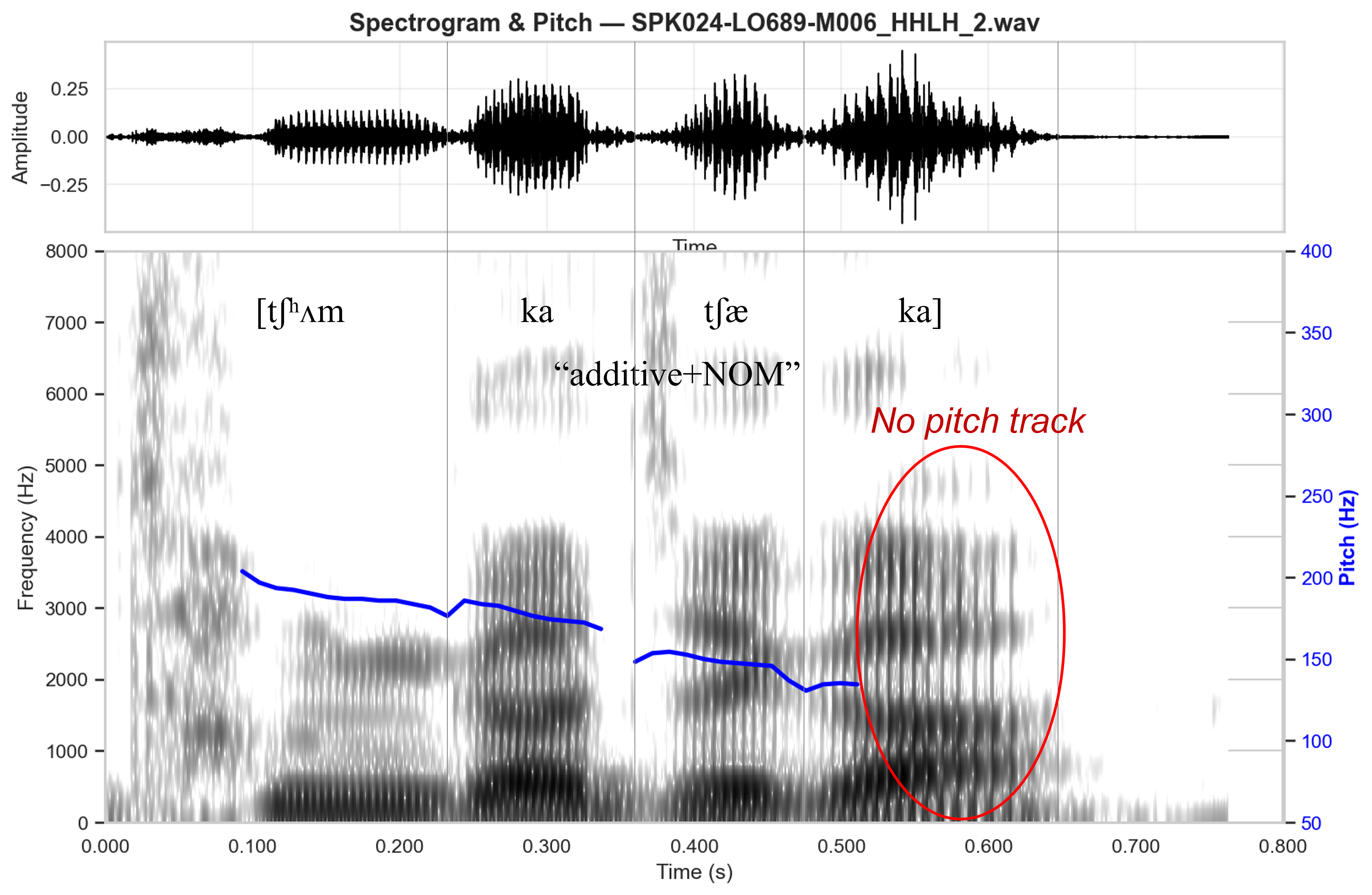}
        \caption{Pitch track loss in the AP-final syllable.}
        \label{fig:a}
    \end{subfigure}
    \hfill
    \begin{subfigure}[b]{0.48\linewidth}
        \centering
        \includegraphics[width=\textwidth, height=4.5cm]{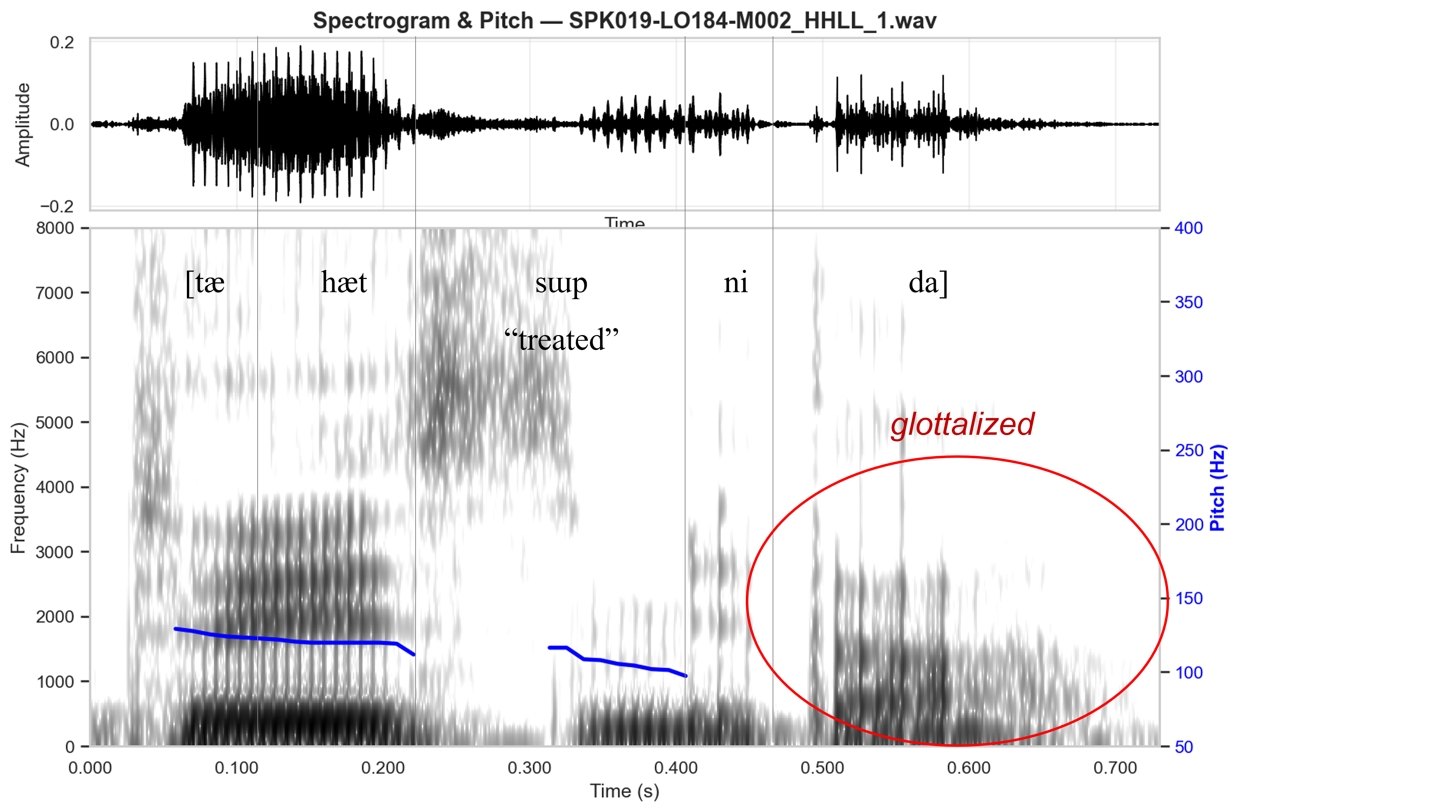}
        \caption{Pitch track loss due to glottalization on the AP-final vowel.}
        \label{fig:b}
    \end{subfigure}

    \vspace{0.5cm}

    \begin{subfigure}[b]{0.48\linewidth}
        \centering
        \includegraphics[width=\textwidth, height=4.5cm]{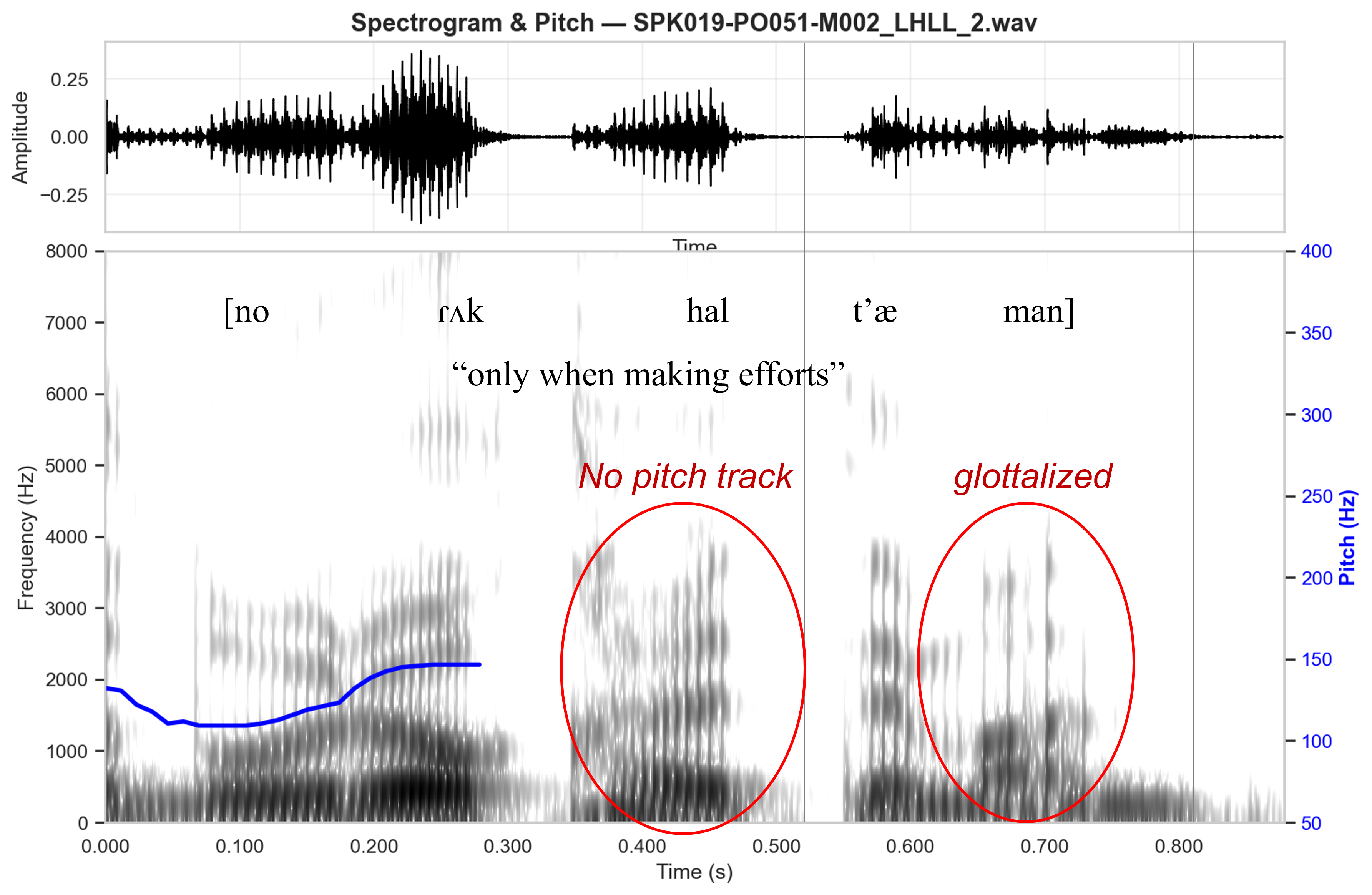}
        \caption{Pitch track loss on the third syllable, as well as on the AP-final vowel due to glottalization.}
        \label{fig:c}
    \end{subfigure}
    \hfill
    \begin{subfigure}[b]{0.48\linewidth}
        \centering
        \includegraphics[width=\textwidth, height=4.5cm]{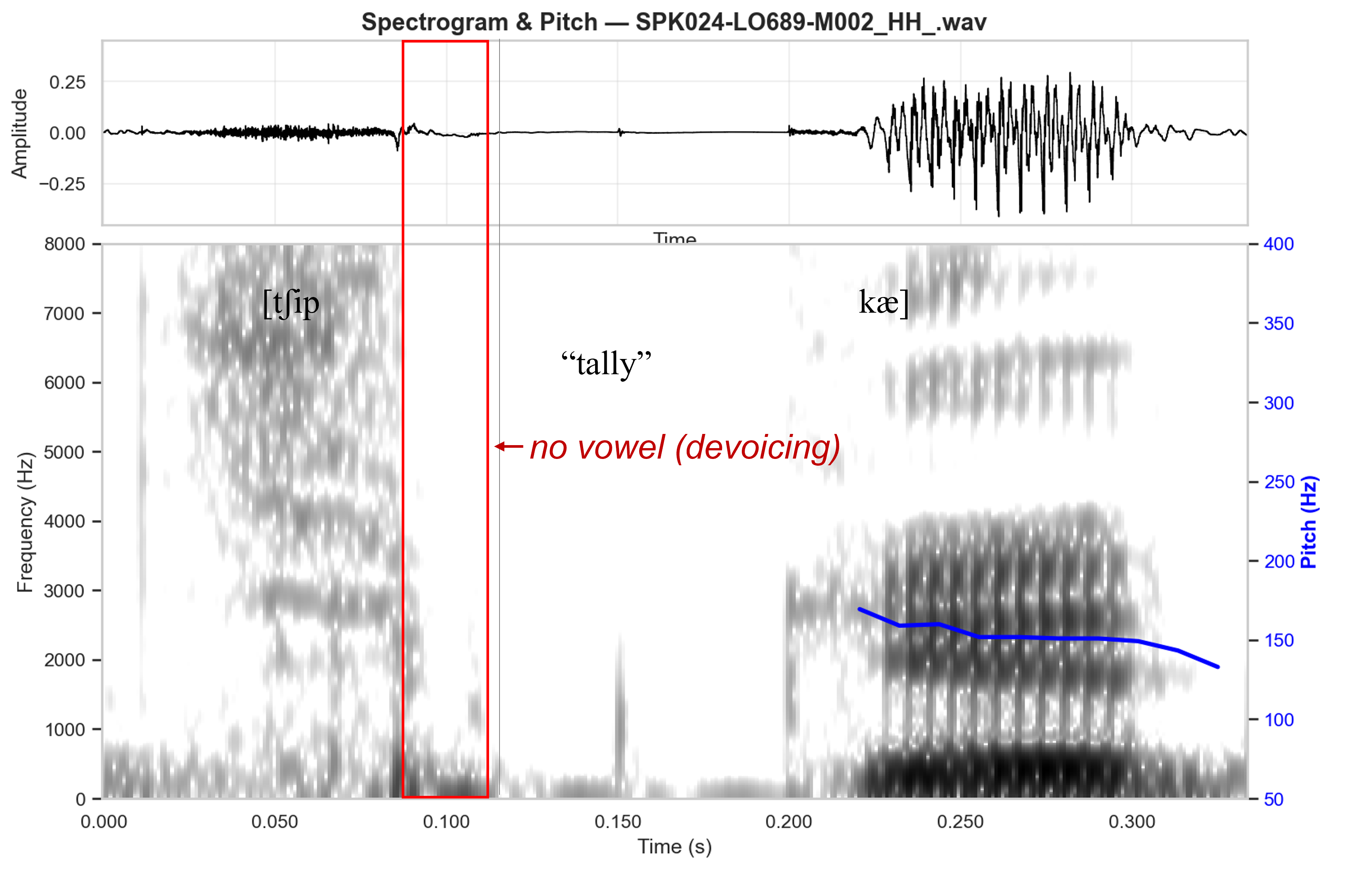}
        \caption{Pitch track loss due to vowel devoicing after [s].}
        \label{fig:d}
    \end{subfigure}

    \vspace{0.5cm}

    \begin{subfigure}[b]{0.48\linewidth}
        \centering
        \includegraphics[width=\textwidth, height=4.5cm]{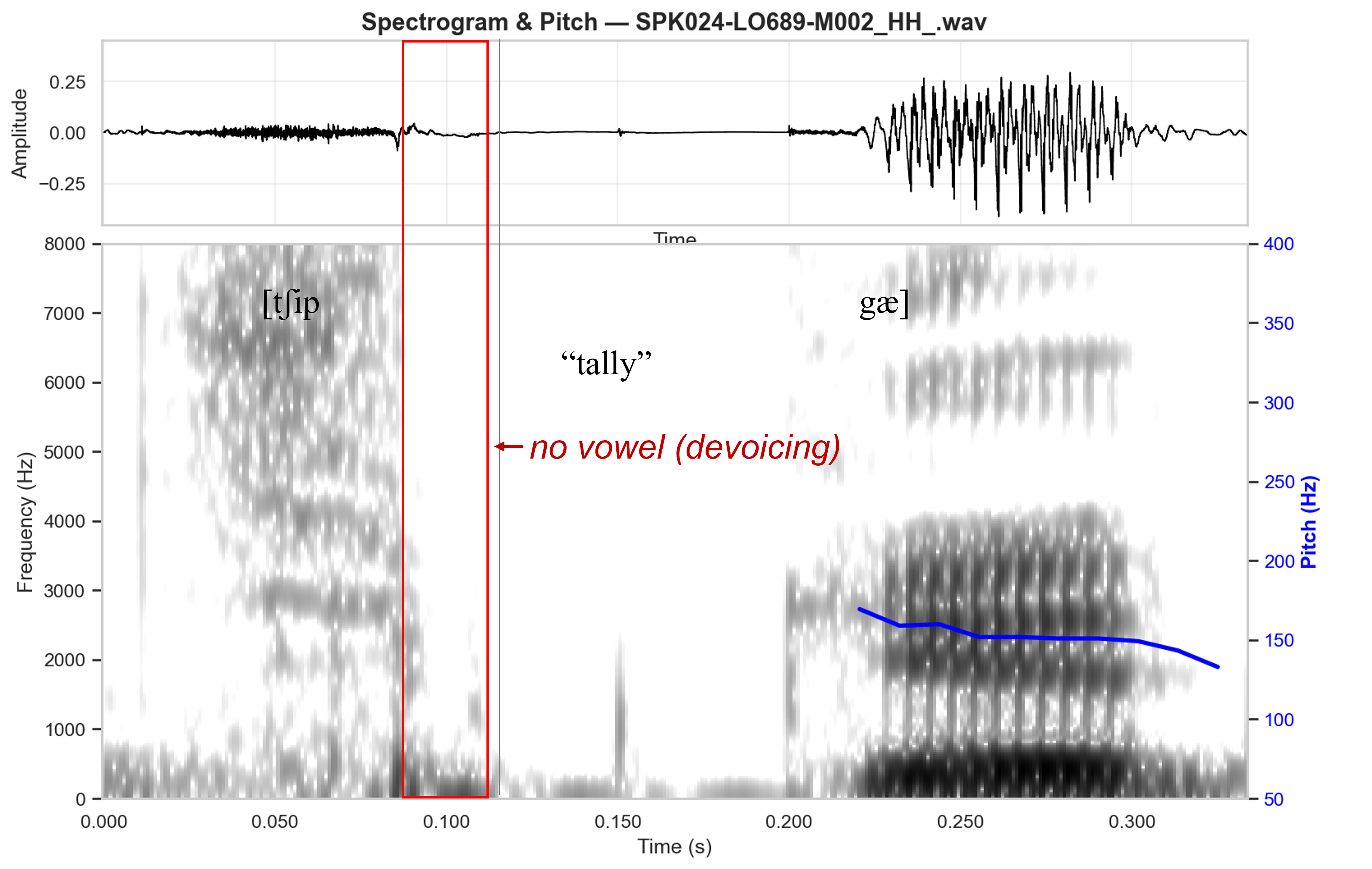}
        \caption{Pitch track loss due to vowel devoicing after the affricate.}
        \label{fig:e}
    \end{subfigure}
    \hfill
    \begin{subfigure}[b]{0.48\linewidth}
        \centering
        \includegraphics[width=\textwidth, height=4.5cm]{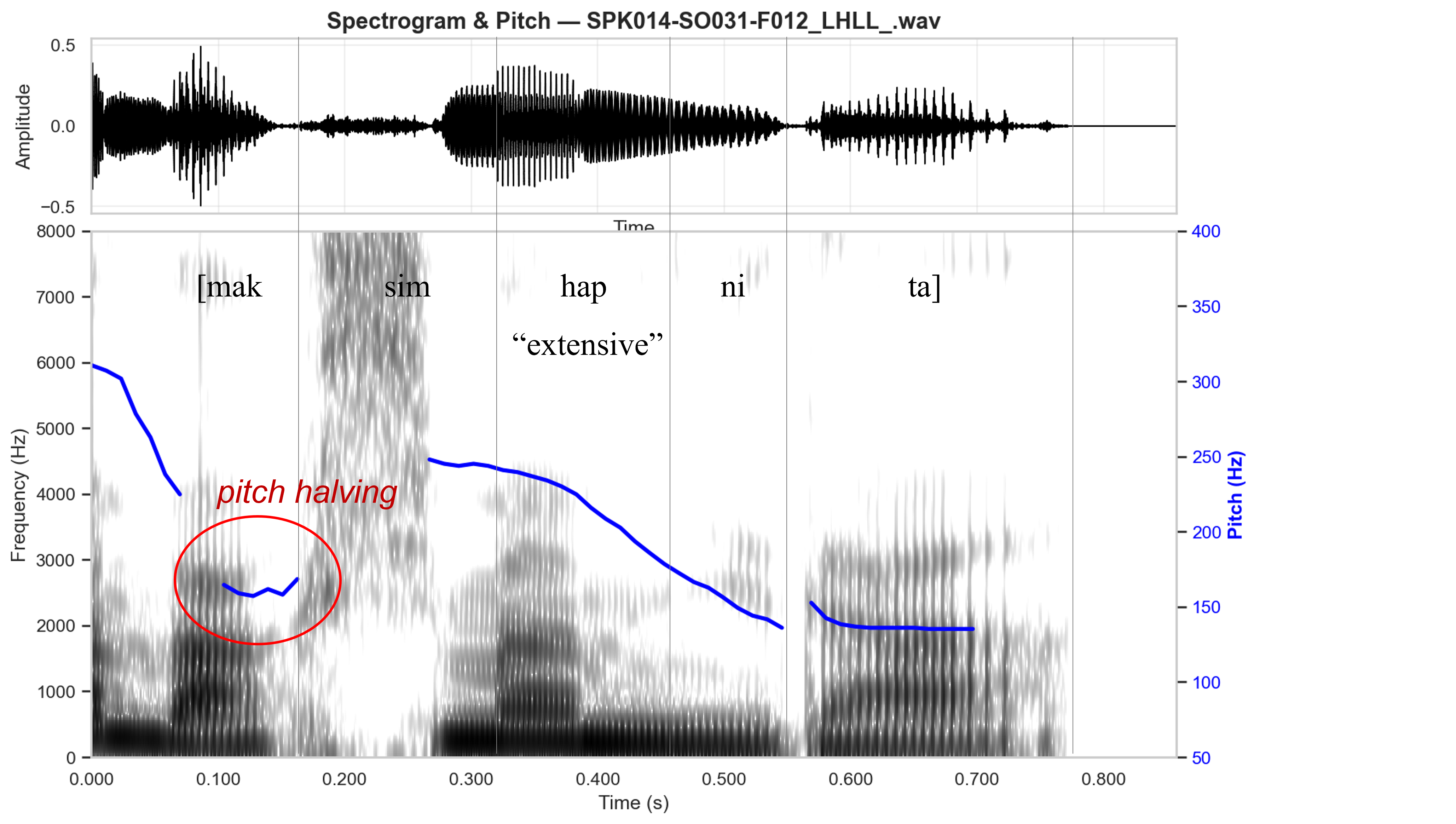}
        \caption{Pitch track error due to pitch halving.}
        \label{fig:f}
    \end{subfigure}

    \vspace{0.5cm}

    \begin{subfigure}[b]{0.48\linewidth}
        \centering
        \includegraphics[width=\textwidth, height=4.5cm]{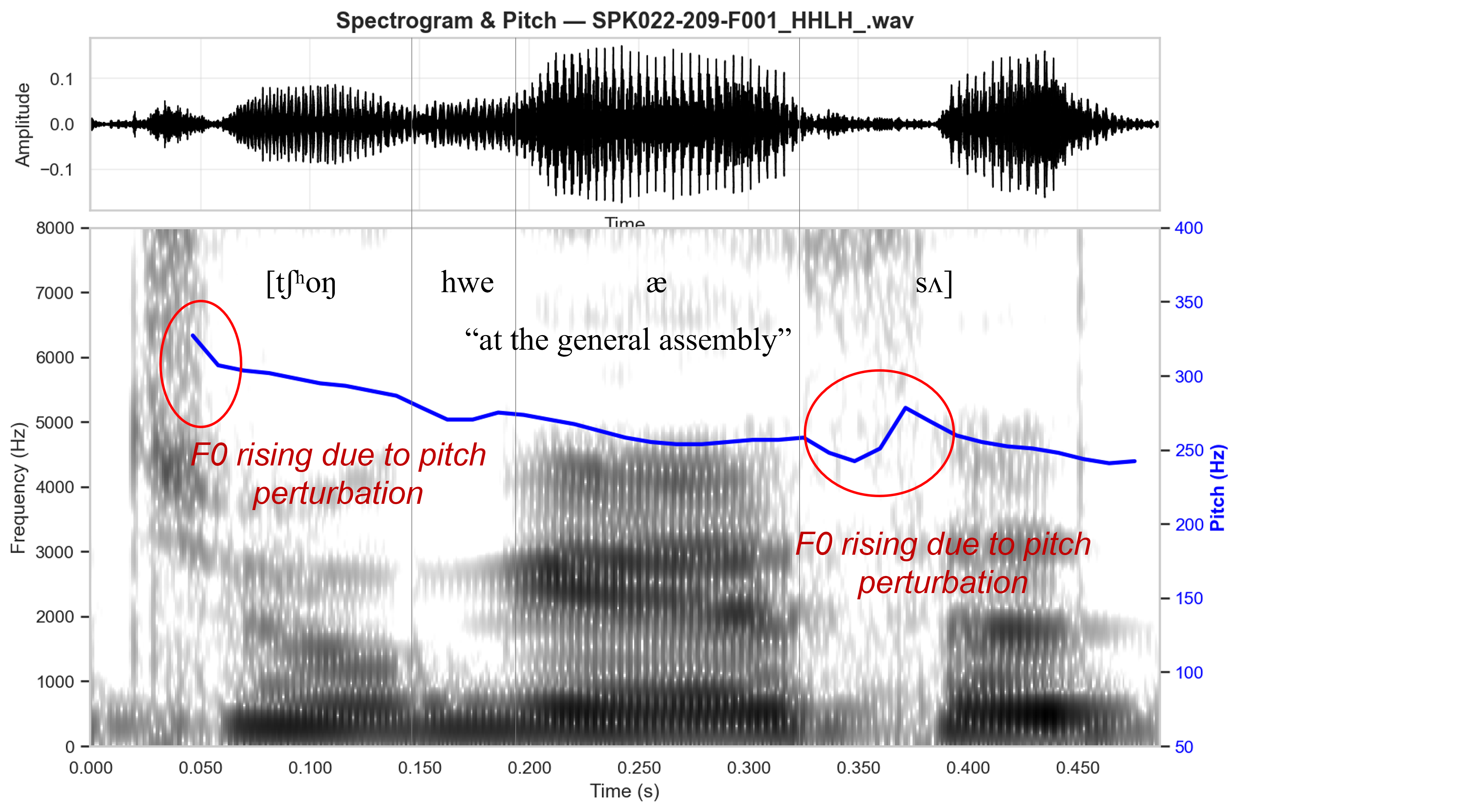}
        \caption{$F_0$ rising due to local pitch perturbation.}
        \label{fig:g}
    \end{subfigure}
    \hfill
    \begin{subfigure}[b]{0.48\linewidth}
        \centering
        \includegraphics[width=\textwidth, height=4.5cm]{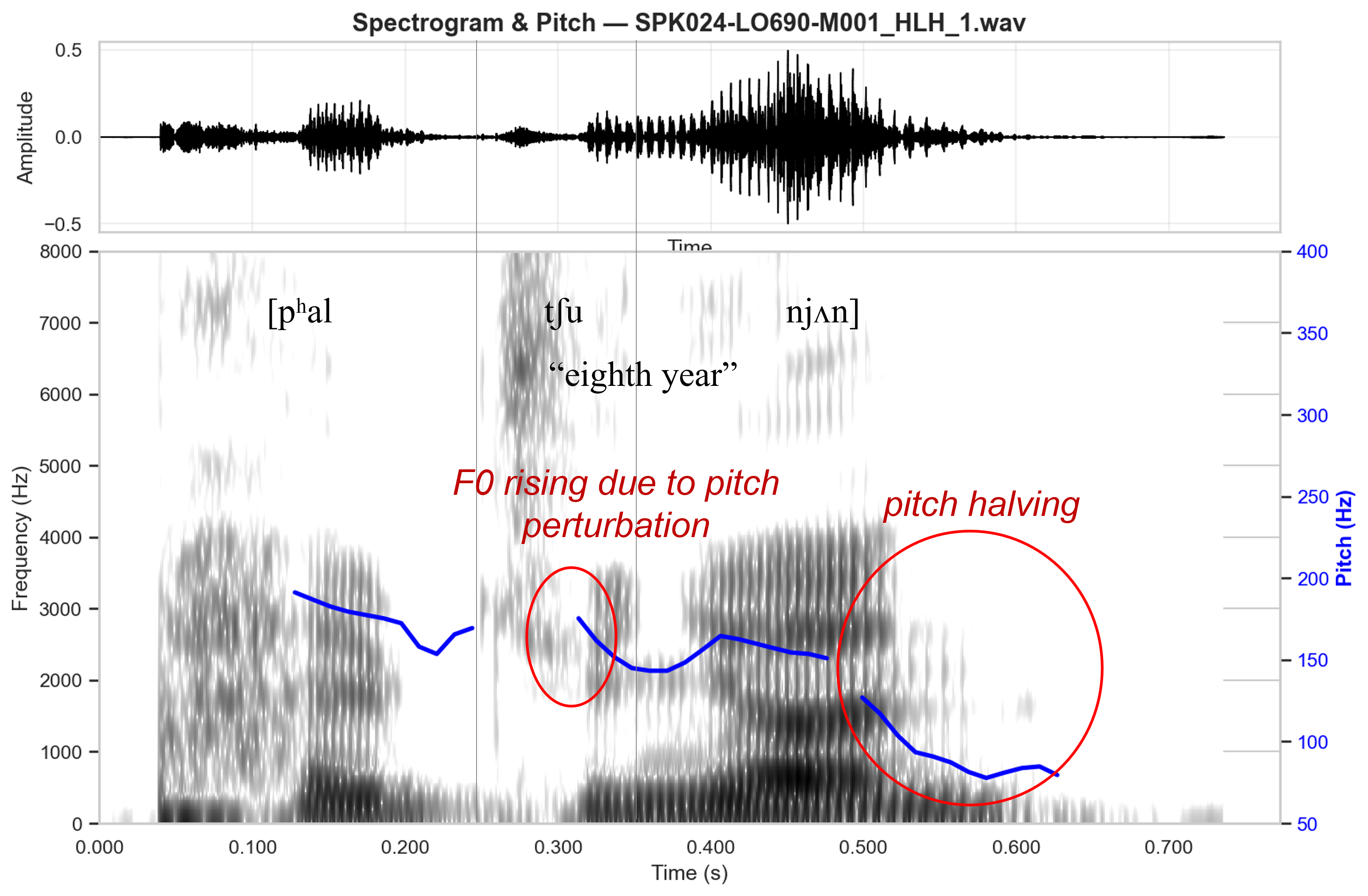}
        \caption{$F_0$ rising due to local pitch perturbation and pitch track error due to pitch halving.}
        \label{fig:h}
    \end{subfigure}

    \vspace{0.5cm}

    \caption{Visual analysis of various $F_0$ discontiunations or pitch track errors in Seoul Korean speech data, including devoicing, pitch halving, glottalization, and $F_0$ perturbation.}
    \label{fig:total_prosody_artifacts}
\end{figure}

The performance of our model is closely related to the quality of the extracted $F_0$ contours. However, as shown in Figure \ref{fig:total_prosody_artifacts}, real-world speech data often includes pitch discontinuations or pitch tracking errors that make precise pitch tracking difficult, leading to potential misclassifications.

First, \text{pitch tracking loss} is quite frequent. The $F_0$ disappears entirely in many cases, such as during vowel devoicing following [s] or the affricate (Figure \ref{fig:d} and \ref{fig:e}) or at sentence-final positions (Figure \ref{fig:a}). These missing portions of the $F_0$ are problematic, as the model loses crucial information required to identify the entire tonal pattern.

Also, pitch tracking errors such as \text{pitch halving} creates sudden drops in the $F_0$ contour (Figure \ref{fig:f} and \ref{fig:h}). When these errors occur, the model may misinterpret a H tone as a L tone. 

Third, the realization of low tones is often accompanied by glottalization on the vowel, in which $F_0$ contours exhibit pitch tracking errors or signal loss. The pattern is particularly frequent in the phrase-final L tones as in LHL\underline{L} or HHL\underline{L} (Figure \ref{fig:b} and \ref{fig:c}).

Lastly, \text{local $F_0$ perturbations} often lead to slight $F_0$ rising following the release of stops and affricates (Figure \ref{fig:g} and \ref{fig:h}).

\newpage

\section*{Appendix G: Effect of Encoder Dimension}
To investigate the impact of the representation capacity on downstream classification performance, we evaluated the extracted features using three classifiers—RF, LR, and LightGBM—across encoder dimensions ranging from 64 to 1024. The results are summarized in Table \ref{tab:encoder_dim}.

Overall, the classification performance exhibited a slight improvement as the encoder dimension increased. In particular, the LR classifier demonstrated the most notable enhancement, achieving the best overall performance of our proposed framework at a dimension of 1024 (Acc: 0.7775, F1: 0.5154). In contrast, while the RF and LightGBM models also showed marginal performance gains, the differences across dimensions were not highly significant. These findings suggest that expanding the encoder dimension generally yields minor performance benefits, but specifically enables the LR model to optimally leverage the high-dimensional latent representations.

\begin{table}[htbp]
\centering
\caption{Performance comparison across different encoder dimensions. All results represent the average of 5-fold cross-validation (Mean $\pm$ SD).}
\label{tab:encoder_dim}
\renewcommand{\arraystretch}{0.75} 
\resizebox{\columnwidth}{!}{%
\begin{tabular}{llccc}
\toprule
\textbf{Encoder Dimension} & \textbf{Metric} & \textbf{RF} & \textbf{LR} & \textbf{LightGBM} \\
\midrule
\multirow{2}{*}{\textbf{64}}
& Acc & 0.7710 $\pm$ 0.0057 & 0.7690 $\pm$ 0.0056 & 0.7694 $\pm$ 0.0062 \\
& F1 & 0.4979 $\pm$ 0.0091 & 0.4701 $\pm$ 0.0148 & 0.4968 $\pm$ 0.0145 \\
\midrule
\multirow{2}{*}{\textbf{128}}
& Acc & 0.7728 $\pm$ 0.0082 & 0.7721 $\pm$ 0.0082 & 0.7686 $\pm$ 0.0087 \\
& F1 & 0.4944 $\pm$ 0.0108 & 0.4880 $\pm$ 0.0106 & 0.4893 $\pm$ 0.0149 \\
\midrule
\multirow{2}{*}{\textbf{256}}
& Acc & 0.7720 $\pm$ 0.0046 & 0.7747 $\pm$ 0.0048 & 0.7731 $\pm$ 0.0065 \\
& F1 & 0.5000 $\pm$ 0.0134 & 0.5029 $\pm$ 0.0110 & 0.5003 $\pm$ 0.0167 \\
\midrule
\multirow{2}{*}{\textbf{512}}
& Acc & 0.7710 $\pm$ 0.0070 & 0.7730 $\pm$ 0.0061 & 0.7689 $\pm$ 0.0083 \\
& F1 & 0.4938 $\pm$ 0.0073 & 0.5004 $\pm$ 0.0133 & 0.4968 $\pm$ 0.0180 \\
\midrule
\multirow{2}{*}{\textbf{1024}}
& Acc & 0.7740 $\pm$ 0.0069 & \textbf{0.7775 $\pm$ 0.0064} & 0.7743 $\pm$ 0.0052 \\
& F1 & 0.5051 $\pm$ 0.0061 & \textbf{0.5154 $\pm$ 0.0157} & 0.5086 $\pm$ 0.0064 \\
\bottomrule
\end{tabular}%
}
\end{table}

\newpage


\end{document}